\documentclass[12pt]{iopart}

\usepackage{graphicx}
\usepackage[compatibility=false]{caption}
\usepackage{subcaption}
\expandafter\let\csname equation*\endcsname\relax
\expandafter\let\csname endequation*\endcsname\relax
\usepackage{amsmath}
\usepackage{amssymb}
\usepackage{bm}
\usepackage{booktabs}
\usepackage{amsfonts}
\usepackage{url}
\usepackage[numbers,sort&compress]{natbib}

\usepackage[dvipsnames]{xcolor}
\colorlet{nonlocal}{ForestGreen} 
\colorlet{loczero}{red}          
\colorlet{zfun}{violet}          
\colorlet{const}{BurntOrange}    

\newcommand{\dbp}{\delta \! B_{\parallel}}

\begin{document}

\title[Global Gyrokinetic Simulations of Electromagnetic Turbulence in STEP]{Global Gyrokinetic Simulations of Electromagnetic Turbulence in STEP}

\author{D. Kennedy$^1$, F. Sheffield$^2,$ T. G{\"o}rler$^2$, C. M. Roach$^1$, M. Giacomin$^{3},$  A. Bokshi$^1$, D. Dickinson$^4$, H. G. Dudding$^{1},$ and B. S. Patel$^{1}$}

\address{$^1$UKAEA (United Kingdom Atomic Energy Authority), Culham Campus, Abingdon,
Oxfordshire, OX14 3DB, UK}
\address{$^2$Max Planck Institute for Plasma Physics, Boltzmannstr. 2, Garching, 85748, Germany}
\address{$^3$Dipartimento di Fisica e Astronomia ``G. Galilei'', Universit\`a degli Studi di Padova, Padova, Italy}
\address{$^4$York Plasma Institute, University of York, York, YO10 5DD, United Kingdom}

\ead{daniel.kennedy@ukaea.uk}
\vspace{10pt}
\begin{indented}
 \item[] 

\end{indented}

\begin{abstract}
This paper presents gradient-driven global electromagnetic gyrokinetic simulations for a conceptual burning flat-top operating point of STEP \cite{tholerus2024}, STEP-EC-HD, and investigates how non-local effects influence the nonlinear saturation and transport of the electromagnetic turbulence at finite~$\beta$. Local gyrokinetic simulations have shown that including $\delta \! B_{\parallel}$ is essential for the dominant hybrid kinetic ballooning modes, or hKBMs, to be unstable in STEP \cite{kennedy2024}. Using the long-wavelength $\delta \! B_{\parallel}$ solver~\cite{sheffield2024} implemented in \texttt{GENE}~\cite{gene}, this work demonstrates that the linear mode spectrum can be accurately captured in global geometry, which results in good agreement with an ensemble of local flux-tube simulations. The global framework reproduces the hKBMs identified in~\cite{kennedy2023a}, while microtearing modes remain challenging to resolve due to their shorter radial scales. Nonlinear simulations reveal clear evidence of an electromagnetic transition to states with extremely large heat fluxes, consistent with local simulations and with the predicted loss of zonal-flow regulation for this proposed operating point~\cite{Kennedy2025EM}. These findings establish the capability of global gyrokinetics to capture finite-$\beta$ dynamics in STEP-like plasmas and motivate future work to identify the conditions governing this transition.

\end{abstract}

\section{Introduction}

The performance of magnetic confinement fusion (MCF) devices such as spherical tokamaks (STs) is often limited by turbulent fluctuations, which are typically the dominant cause of heat, particle, and momentum transport losses. Accurately predicting the transport driven by these fluctuations is essential for the optimisation of future STs. The UK STEP programme~\cite{STEP,meyer2022,meyer2024} aims to produce more than 100~MW of net electric power by developing a compact power plant based on the ST concept. In its initial phase, the programme has focused on designing reference plasma equilibria using simplified empirical transport models guided by confinement scalings~\cite{tholerus2024}, which have been successful in predicting turbulence in modern conventional-aspect-ratio devices~\cite{tglf,qualikiz,citrin2022}. However, next-generation tokamaks such as STEP are expected to operate at higher values of plasma $\beta$ [where $\beta \equiv 2\mu_0 p / B^2$ is the ratio of the plasma pressure $p$ to the magnetic field energy density $B^2/(2\mu_0)$, with $B$ the magnetic field strength and $\mu_0$ the permeability of free space] than present-day conventional-aspect-ratio tokamaks. As a result, STs are expected to show a stronger electromagnetic character in their transport. In these conditions, first-principles models such as gyrokinetics (GK) are required to provide more accurate predictions.

The first GK analysis of one of STEP's SPP-1 reference flat-top operating points, STEP-EC-HD~\cite{tholerus2024}, was conducted in~\cite{kennedy2023a}. This study identified unstable \textit{hybrid} kinetic ballooning modes (hKBMs) as the dominant instabilities, with subdominant microtearing modes (MTMs) also present; the hKBMs combine features of kinetic ballooning modes (KBMs), ion-temperature-gradient-driven modes (ITGs), and trapped-electron modes (TEMs)\footnote{Note that~\cite{giacomin2023b} reports the results of local GK calculations for a single flux-surface close to mid-radius. Deeper into the core MTMs can become the dominant instability~\cite{giacomin2024b}.}. Subsequent nonlinear local simulations~\cite{giacomin2023b,kennedy2024} revealed that, in the absence of equilibrium flow shear, hKBM-driven turbulence can generate particle and heat fluxes far exceeding the available heating and fuelling rates. Simulations of STEP's reference flat-top operating points are often difficult to saturate, with phases of apparent saturation punctuated by transient bursts of transport and a transition to extreme heat fluxes. However, it was also shown in \cite{giacomin2023b} that, in scans where $\beta^{\prime} \propto \beta$, saturation at more moderate fluxes is possible at both larger and smaller values of $\beta,$ or even at the nominal value of $\beta$ by including modest amounts of equilibrium flow shear (though with no torque from NBI in STEP-EC-HD, the rotation is expected to be modest) in the simulation. A first flux-driven transport prediction for a STEP flat top~\cite{giacomin2024b}, based on a new reduced model for turbulent transport from hKBMs and other simplifying assumptions, finds a transport steady state (enabled by $\beta^\prime$ stabilisation~\cite{bourdelle2003}) with fusion power comparable to that assumed in JETTO, a simplified transport code, although it did not demonstrate how such a state might be accessed~\cite{giacomin2024b}.

The nonlinear simulations in~\cite{giacomin2023b,kennedy2024} employed the local (flux-tube) approximation~\cite{beer1995}, which simplifies the study of plasma turbulence by assuming a clear scale separation between small-scale fluctuations and large-scale equilibrium variations. The computational domain is assumed to be confined to a narrow region around a single magnetic flux surface, with turbulence radially localised in this domain and equilibrium gradients treated as constant across the domain. In experiments, however, this separation of scales is often imperfect: turbulence spans a broad range of spatial and temporal scales, and the turbulent perpendicular length-scale can approach equilibrium scales.  Such multiscale dynamics, along with intrinsically global effects like profile shearing, is not captured within the strictly local framework. Global gyrokinetic simulations overcome these limitations by incorporating the full radial variation of equilibrium quantities, allowing self-consistent treatment of non-local phenomena such as profile curvature, turbulent avalanches, internal transport barriers, and turbulence spreading (see e.g.,~\cite{garbet2010} and references therein).

The very large fluxes predicted by local simulations in the absence of perpendicular shear flows (which can suppress the largest scale)~\cite{giacomin2023b,kennedy2024,Kennedy2025EM} raise the question of whether effects omitted in the flux-tube approximation could contribute to this discrepancy. In particular, profile shearing and coupling to radially extended mode structures (global effects neglected in the local approach) may influence both the linear drive and the nonlinear balance between turbulence and zonal flows, the principal regulators of ion-scale transport~\cite{Kennedy2025EM}. By retaining finite-$\rho_\star$ variations of equilibrium quantities and enabling self-consistent interaction with evolving radial profiles, global simulations provide a means to assess whether the large fluxes predicted locally persist once global physics is included. Indeed, recent global electromagnetic simulations in stellarator geometries using \texttt{EUTERPE} and \texttt{GENE-3D} suggest that such effects can qualitatively alter turbulence characteristics and lead to more moderate fluxes than inferred from local studies~\cite{Mishchenko2023,Wilms2024NF}. 

Given that accurate transport predictions are central to assessing the viability of STEP operating scenarios, exploring the impact of global physics on predicted transport levels is therefore of critical importance. This paper reports a first step in this direction, presenting the first gradient-driven electromagnetic global GK simulations of STEP, using the \texttt{GENE} code, to assess the impact of global effects on the predicted level of turbulent transport.
Global gyrokinetic simulations remain computationally demanding and technically complex, requiring \emph{ad hoc} boundary conditions, careful formulation of field equations to avoid unphysical sources and reduced numerical efficiency. Until recently~\cite{sheffield2024,wilms2024}, most global codes could not include full electromagnetic physics: in particular, parallel magnetic fluctuations ($\delta\! B_\parallel$), shown in~\cite{kennedy2024} to be essential for hKBM instability in STEP-EC-HD, were absent from the global version of \texttt{GENE}~\cite{GENEglobal} and only available in a few particle-in-cell codes~\cite{Mishchenko2023,yang2023}. An arbitrary-wavelength $\dbp$ solver has now been implemented in global \texttt{GENE}, enabling, in principle, fully electromagnetic nonlinear simulations, though so far applied only to simplified cases. A more recent long-wavelength electromagnetic treatment~\cite{sheffield2024}, in which gyroaveraging in the \(\dbp\) equation is neglected at leading order under the assumption \(k_\perp \rho_i \ll 1\), significantly reduces the computational cost and has opened the way to global, fully electromagnetic turbulence studies of STEP-like plasmas, which are the focus of this work. Here \( \rho_i \) is the ion thermal gyroradius and \( k_\perp \) is the perpendicular wavenumber with respect to the equilibrium magnetic field. This approximation is well suited to the hKBM regime relevant to STEP, as the dominant electromagnetic fluctuations are pressure-gradient-driven and occur at ion-scale perpendicular wavelengths with \(k_\perp \rho_i \ll O(1)\).

In this paper, we present the first global electromagnetic gyrokinetic simulations of the STEP-EC-HD operating point. Building on previous local studies~\cite{giacomin2023b,kennedy2024}, our aim is to determine whether the extremely large fluxes reported in flux-tube simulations persist once global profile variation and magnetic geometry are included. By exploiting recent advances in the global version of \texttt{GENE}~\cite{sheffield2024}, we carry out nonlinear gradient-driven simulations that retain the essential electromagnetic physics, including $\delta \! B_\parallel$, within a fully global gradient-driven framework. These simulations provide the first test of whether global effects, such as finite-$\rho_\star$ profile variation and the coupling to radially extended mode structures, can reduce hKBM turbulence and bring transport fluxes closer to the available sources in STEP-EC-HD. 

The layout of this paper is as follows. Section~\ref{sec:global_GK_intro} discusses global GK, introducing the equations solved by global \texttt{GENE} and addressing the physics basis for including global effects in GK simulations. Section~\ref{sec:STEP-equilibrium} introduces the STEP equilibrium used in this work and the numerical setup. Sections~\ref{sec:linear} and~\ref{sec:nonlinear} show the results of linear and nonlinear GK simulations comparing both global and local approaches. 
Finally, our conclusions are presented in Section~\ref{sec:conclusions}. 

\section{Global electromagnetic $\delta \! f$ gyrokinetics} \label{sec:global_GK_intro}

In this section, we remind the reader of precisely what is meant by global $\delta \! f$ gyrokinetics; and reintroduce the equations that are being solved by \texttt{GENE} (see e.g., \cite{sheffield2024,wilms2024} for further detail) before discussing the need for global GK. This material establishes the notation, the local-to-global hierarchy, and the modelling choices (in particular the treatment of $\delta \! B_\parallel$ and of gradient-driven sources) on which the results of Sections~\ref{sec:linear} and~\ref{sec:nonlinear} rely; readers already familiar with global $\delta\!f$ gyrokinetics may prefer to skip Section~\ref{sec:global_GK_intro} and proceed directly to Section~\ref{sec:STEP-equilibrium}.

\subsection{Hierarchy of gyrokinetic formalisms}

Gyrokinetic simulations can be classified along several axes, summarised in Table~\ref{tab:GK_classification}. 
\textit{Local} (flux-tube) simulations neglect radial equilibrium variation and typically adopt a domain that is periodic and narrow ($\sim$ few Larmor radii wide) in directions $\perp \mathbf{B}$, and highly extended $\parallel \mathbf{B}$: i.e. a domain that is suitable for long ($\parallel \mathbf{B}$) thin eddies. 
\textit{Global} simulations, on the other hand, retain radial profile variation, and adopt a larger global domain that can capture mesoscale and non-local effects and thus provide a more physically comprehensive description. 
\textit{Gradient-driven} simulations fix background gradients (e.g.\ the linear drive 
terms in the GK equation), 
while \textit{flux-driven} simulations evolve the profiles self-consistently in the presence of prescribed sources and sinks. 
The \textit{$\delta\! f$} approach evolves only the fluctuating part of the distribution function around a fixed equilibrium, 
in contrast to the more general \textit{full-$F$} approach, which evolves the entire distribution and is required when the equilibrium changes significantly and typically requires confinement time scale simulations. Finally, simulations may adopt an \textit{Eulerian} formulation, in which the distribution function is discretised on a fixed phase-space grid, 
or a \textit{Lagrangian} formulation (often particle-in-cell), in which the distribution is represented by markers following phase-space trajectories.

\begin{table}[h]
\caption{Classification of gyrokinetic simulations along several common axes.}
\centering
\begin{tabular}{lll}
\hline
\textbf{Axis} & \textbf{Option} & \textbf{Summary} \\
\hline
Radial extent & Local (flux-tube) & Narrow domain; equ'm/turbulence radial scale separation\\
              & Global            & Retains profiles; captures mesoscale/non-local effects \\
\hline
Profile evolution & Gradient-driven & Fixed background gradients \\
                  & Flux-driven     & Profiles evolve with sources/sinks \\
\hline
Distribution & $\delta \!f$   & Fluctuations about fixed equilibrium \\
             & Full-$F$     & Total distribution, equilibrium included \\
\hline
Formulation  & Eulerian     & Fixed phase-space grid (continuum codes) \\
             & Lagrangian   & Field-line/particle following (PIC methods) \\
\hline
\end{tabular}
\label{tab:GK_classification}
\end{table}

These distinctions form a natural hierarchy of models~\cite{2010NucFu..50d3002G}: local, gradient-driven, $\delta \! f$, Eulerian simulations are the most efficient and widely used for parameter scans, 
while global, flux-driven, full-$F$, Lagrangian simulations are the most complete but also the most computationally demanding. 
In practice, different levels in this hierarchy are employed depending on the physics questions of interest. 
For STEP-EC-HD, gyrokinetic studies to date~\cite{kennedy2023a,Kennedy2025EM,giacomin2023b,kennedy2024} 
have mainly been restricted to local, gradient-driven calculations, with the exception of~\cite{giacomin2024b},
which reports flux-driven simulations using a quasilinear model for the nonlinear fluxes. 
In this work we advance one step further up the hierarchy by carrying out the first global, gradient-driven gyrokinetic simulations of STEP-EC-HD,
undertaken to capture mesoscale and non-local effects that are inaccessible to local approaches. By \emph{mesoscale} we mean structures on scales intermediate between the gyroradius and the minor radius, $\rho_i \ll \lambda \ll a$, such as radially extended eigenmodes spanning many rational surfaces, turbulence spreading, avalanches, and zonal profile corrugations; these are ordered out of the local approximation but do not require full device-scale evolution.

\subsection{Gradient-driven global simulations}

We are interested in plasmas that are well described by the GK framework (see, e.g.,~\cite{Abel2013,Catto2019}). Specifically, we consider fluctuations proportional to $\sim \exp[ \mathrm{i} ( \mathbf{k} \cdot \mathbf{r} -  \omega t ) ]$, with frequency $\omega$, parallel and perpendicular wavenumbers $k_\parallel$ and $k_\perp$ relative to the equilibrium magnetic field direction $\mathbf{b}_{0} = \mathbf{B}_{0}/B_{0}$, that satisfy the standard GK orderings
\begin{equation}
\frac{\omega}{\Omega_{\sigma}} \sim \frac{\nu_{\sigma\sigma^{\prime}}}{\Omega_{\sigma}} \sim \frac{k_\parallel}{k_\perp} \sim \frac{q_{\sigma}\,\delta\phi}{T_{0\sigma}} \sim \frac{\delta \!B_\parallel}{B_{0}} \sim \frac{|\delta \mathbf{B}_{\perp}|}{B_{0}} \sim \rho_{\star,\sigma} \ll 1, \label{eq:gk_ordering}
\end{equation}
where $\Omega_\sigma \equiv q_\sigma B_0/(m_\sigma c)$ is the gyrofrequency, $q_{\sigma}$ and $m_{\sigma}$ are charge and mass of species $\sigma$, $\nu_{\sigma\sigma'}$ is the collision frequency between species $\sigma$ and $\sigma'$. The expansion parameter, $\rho_{\star,\sigma} \equiv \rho_\sigma/L,$ is the ratio of the Larmor radius $\rho_\sigma \equiv v_{\mathrm{th},\sigma} / \Omega_\sigma$ to a macroscopic equilibrium scale length $L$, $\phi$ is the electrostatic potential, and $\delta  \! B_\parallel$ and $\delta \mathbf{B}_\perp$ are the parallel and perpendicular components of the perturbed magnetic field. In the global $\delta\!  f$ formulation, $L$ is allowed to be comparable to the radial extent of the computational domain, and no assumption of a local (flux-tube) expansion is made (see Section~\ref{subsec:local_vs_global}). We define the standard gyroaverages (average over the gyroangle $\vartheta$) at constant real position $\bm{r}$ and guiding-centre position $\bm{R}_{\sigma} = \bm{r} - \mathbf{b} \times \bm{v}_{\perp} /\Omega_{\sigma}$, through the operators
\begin{equation}
\mathcal{K}\{u\}(\bm{r}) \equiv 
\frac{1}{2\pi}\int_{0}^{2\pi}
u\left(\bm{r}-\boldsymbol{\rho}(\vartheta),\bm{v},t\right)\,\mathrm{d}\vartheta,
\end{equation}
\begin{equation}
\mathcal{G}\{u\}(\bm{R}_\sigma)
\;\equiv\;
\frac{1}{2\pi}\int_{0}^{2\pi}
u\left(\bm{R}_\sigma+\boldsymbol{\rho}(\vartheta),\bm{v},t\right)\,\mathrm{d}\vartheta.
\end{equation}

The gyrocenter distribution, $F_\sigma(\bm{R}_\sigma,\bm{v},t)\equiv G\{f_\sigma\}(\bm{R}_\sigma,\bm{v},t),$ where $f_{\sigma}$ is the particle distribution function, evolves according to the full-$F$ gyrocenter Vlasov-Landau equation (e.g.,~\cite{garbet2010}):
\begin{equation}
\frac{\partial F_\sigma}{\partial t}
+ \dot{\bm{R}}\cdot\nabla F_\sigma
+ \dot{v}_\parallel\frac{\partial F_\sigma}{\partial v_\parallel}
= C_\sigma[F_\sigma] + S_\sigma, \label{eq:Vlasov_full-F}
\end{equation}
where $C_\sigma[\cdot]$ is the collision operator for species $\sigma$ and $S_\sigma$ is a generic source term for species $\sigma$. \texttt{GENE} (and a majority of other GK codes) solve this equation in the GK ordering~(\ref{eq:gk_ordering}) by appealing to a technique known as  $\delta \! f$ splitting where the full gyrocenter distribution function $F_{\sigma}$ is split into a stationary background distribution function $F_{0,\sigma}$, describing the plasma at thermal equilibrium, and a time-dependent, first-order perturbation $F_{1,\sigma}$:

\begin{equation}
F_\sigma = F_{0,\sigma} + F_{1,\sigma}, \quad \frac{\|F_{1,\sigma}\|}{\|F_{0,\sigma}\|} \sim O(\epsilon_\delta) \ll 1,
\end{equation}
where \( \epsilon_\delta \) is a small parameter which contains the assumption that fluctuations in the system are much weaker (i.e., have asymptotically smaller amplitude) than the stable background profile. The parameter \( \epsilon_\delta \)  is taken to be such that $\epsilon_\delta \sim \rho_{\star \sigma}.$ 
In this work we assume that the background distribution is given by a local Maxwellian:
\begin{equation}
F_{0,\sigma} = F_{M,\sigma}(x, z, v_{\parallel}, \mu) = \frac{n_{0,\sigma}(x)}{\pi^{3/2} v_{\text{th},\sigma}^3(x)} \exp \left( - \frac{m_\sigma v_{\parallel}^2/2 + \mu B_0(x, z)}{T_{0,\sigma}(x)} \right)
\end{equation}
where $m_\sigma$, $n_{0,\sigma}$ and $T_{0,\sigma}$ are the mass, equilibrium background density, and temperature profiles of species $s$ respectively. Phase space is discretised in $(x,y,z,v_\parallel,\mu)$, where $y$ is the binormal coordinate (perpendicular to both $x$ and the equilibrium magnetic field), and $z$ the coordinate along the field line. The velocity-space variables are the parallel velocity $v_\parallel$ and the magnetic moment $\mu = m v_\perp^{2}/(2B_0)$, where $m$ is the particle mass, $v_\perp$ the perpendicular velocity, and $B_0$ the equilibrium magnetic field strength. The thermal velocity $v_{\text{th},\sigma}$ is given by
\begin{equation}
v_{\text{th},\sigma}(x) = \sqrt{\frac{2 T_{0,\sigma}(x)}{m_\sigma}}.
\end{equation}
The magnetic moment, $\mu \equiv m_s v_\perp^2/(2B_0),$ and the parallel velocity, $v_{\parallel}$ are used as velocity-space coordinates in \texttt{GENE}.  

The full global $\delta F$ gyrokinetic equation can then be written as~\cite{wilms2024}
\begin{align}
\frac{\partial F_{1,\sigma}}{\partial t}
&= -\Big[ v_{\parallel}\mathbf{b} + \big(\mathbf{v}_\chi + \mathbf{v}_{\nabla B} + \mathbf{v}_c\big) \Big]\!\cdot\!\nabla F_{1,\sigma}
+ \frac{\mu}{m_\sigma}\,\mathbf{b}\!\cdot\!\nabla B_0\,\frac{\partial F_{1,\sigma}}{\partial v_{\parallel}}
\nonumber\\
&\quad
-\,\mathbf{v}_\chi\!\cdot\!\Bigg[
\nabla\ln n_{0,\sigma}
+ \nabla\ln T_{0,\sigma}\!\left(\frac{m_\sigma v_{\parallel}^2}{2T_{0,\sigma}} + \frac{\mu B_0}{T_{0,\sigma}} - \frac{3}{2}\right)
\Bigg]F_{M,\sigma}
\nonumber\\
&\quad
-\,\frac{q_\sigma F_{M,\sigma}}{T_{0,\sigma}}
\Big[ v_{\parallel}\,\mathbf{b} + \big(\mathbf{v}_\chi + \mathbf{v}_{\nabla B} + \mathbf{v}_c\big) \Big]\!\cdot\!\nabla\,\mathcal{G}\{\psi_1\}
- \frac{q_\sigma v_{\parallel}}{c}\,\frac{F_{M,\sigma}}{T_{0,\sigma}}\,
\frac{\partial}{\partial t}\,\mathcal{G}\{A_{\parallel}\} \nonumber\\ &\quad +C_{\sigma}[F_{1,\sigma}] + S_{\sigma}, \label{eq:GK_equation}
\end{align}
where we have introduced modified GK potentials 
\begin{equation}
\psi_1 \equiv \phi -\frac{\mathbf{v}_\perp\!\cdot\!\mathbf{A}_{1\perp}}{c},\quad
 \chi_1 \equiv \psi_1 - \frac{v_{\parallel}}{c}A_{\parallel}. \label{eq:GK_potentials} \end{equation}
In Equation~\ref{eq:GK_potentials}, the vector potential parallel to the magnetic field is represented by $A_{\parallel}$ (with $A_{\parallel}\equiv \bm{A}_1 \cdot \mathbf{b}_{0}$), $\phi$ is the fluctuating electrostatic potential and $\bm{A}_{1\perp}$ is the perpendicular component of the vector potential [$\delta \! B_\parallel
= - \nabla_\perp \cdot ({\mathbf{b}}_0 \times \mathbf{A}_{1\perp})
$]. Equation~(\ref{eq:GK_equation}) also introduces the generalised $E\times B$ velocity, the $\nabla B$-drift velocity, and the curvature-drift velocity
\begin{align}
\mathbf{v}_\chi
&\equiv \frac{c}{B_0^{2}}\mathbf{B}_0 \times \nabla\,\mathcal{G}\{\chi_1\},\\
\mathbf{v}_{\nabla B}
&\equiv \frac{\mu c}{q_\sigma B_0^{2}}\mathbf{B}_0 \times \nabla B_0,\\
\mathbf{v}_c
&\equiv \frac{v_{\parallel}^2}{\Omega_\sigma}\mathbf{b}\times\left(\nabla\ln B_0 + \frac{\beta}{2}\nabla\ln p_0\right).
\end{align}

Equation~(\ref{eq:GK_equation}) is closed via Maxwell's equations, which in this notation read
\begin{align}
-\,\nabla_\perp^2 \phi
&+ \sum_\sigma q_\sigma^2 \int
\Bigg[
\left(\frac{F_{M,\sigma}}{T_{0,\sigma}}\right)\phi
- \mathcal{K}\!\left\{\frac{F_{M,\sigma}}{T_{0,\sigma}}\,\mathcal{G}\{\phi\}\right\}
\nonumber\\
&\qquad\qquad\qquad\qquad
+ \mathcal{K}\!\left\{\frac{F_{M,\sigma}}{c\,T_{0,\sigma}}\,\mathcal{G}\{\mathbf{v}_\perp\!\cdot\!\mathbf{A}_{1,\perp}\}\right\}
\Bigg]\,\mathrm{d}^3 \bm{v}
= \sum_\sigma q_\sigma \int \mathcal{K}\{F_{1,\sigma}\}\,\mathrm{d}^3 \bm{v}. \label{eq:Maxwell1}
\end{align}
\begin{equation}
\nabla_\perp^2 A_{\parallel}
= -\,\frac{4\pi}{c}\sum_\sigma q_\sigma \int v_{\parallel}\,\mathcal{K}\{F_{1,\sigma}\}\,\mathrm{d}^3 \bm{v}.\label{eq:Maxwell2}
\end{equation}
\begin{align}
\big[\nabla_\perp \times \mathbf{B}_1\big]_\perp
&= \frac{4\pi}{c}\sum_\sigma \int \Bigg(
q_\sigma\,\mathcal{K}\{\mathbf{v}_\perp F_{1,\sigma}\}
+ q_\sigma^2\,\mathcal{K}\!\left\{\frac{F_{M,\sigma}}{T_{0,\sigma}}\,
\mathbf{v}_\perp\,\mathcal{G}\{\phi\}\right\}
\nonumber\\
&\qquad\qquad\qquad\qquad
- q_\sigma^2\,\mathcal{K}\!\left\{\frac{F_{M,\sigma}}{c\,T_{0,\sigma}}\,
\mathbf{v}_\perp\,\mathcal{G}\{\mathbf{v}_\perp\!\cdot\!\mathbf{A}_{1,\perp}\}\right\}
\Bigg)\,\mathrm{d}^3 \bm{v} . \label{eq:Maxwell3}
\end{align}

\subsubsection{Long-wavelength approximation}

Equation~\eqref{eq:Maxwell3} represents the general gyrokinetic form of the perpendicular Amp\`ere's law. In the long-wavelength limit \(k_\perp \rho_i \ll 1\), the gyroaveraging operators \(\mathcal{G}\) and \(\mathcal{K}\) may be neglected at leading order, such that \(\mathcal{G}\{\cdot\} \simeq \cdot\) and \(\mathcal{K}\{\cdot\} \simeq \cdot\). Under this approximation, finite-Larmor-radius effects are suppressed and the electromagnetic closure for \(\dbp\) simplifies considerably~\cite{sheffield2024} to give
\begin{align}
\big[\nabla_\perp \times \mathbf{B}_1\big]_\perp
&\simeq \frac{4\pi}{c}\sum_\sigma \int \Bigg(
q_\sigma\,\mathbf{v}_\perp F_{1,\sigma}
+ q_\sigma^2\,\frac{F_{M,\sigma}}{T_{0,\sigma}}\,
\mathbf{v}_\perp\,\phi
- q_\sigma^2\,\frac{F_{M,\sigma}}{c\,T_{0,\sigma}}\,
\mathbf{v}_\perp\,(\mathbf{v}_\perp\!\cdot\!\mathbf{A}_{1,\perp})
\Bigg)\,\mathrm{d}^3 \bm{v}.
\label{eq:Maxwell3_LW}
\end{align}
\subsection{On local gyrokinetics versus global gyrokinetics} \label{subsec:local_vs_global}

In this work we use \texttt{GENE} to solve Equations (\ref{eq:GK_equation}), (\ref{eq:Maxwell1}), (\ref{eq:Maxwell2}), and (\ref{eq:Maxwell3_LW}). The formulation of the problem (both conceptually and numerically) changes depending on whether or not the local approximation is used. 

\subsubsection{Local formulation of the GK equations}

In the local, or flux-tube, approximation, all equilibrium quantities, such as density, temperature, equilibrium gradient scale lengths, and magnetic geometry, are evaluated on a single reference flux surface and treated as constant across the perpendicular domain, with only the lowest-order radial gradients retained through a Taylor expansion~\cite{beer1995}. This approximation corresponds to the limit in which the radial extent of the computational domain is asymptotically small compared to the equilibrium scale lengths [i.e.\ the \(L \rightarrow \infty\) limit of (\ref{eq:GK_equation})], leading to a clear separation between the turbulence scales and the equilibrium variation.

A key consequence of this approximation is that the perpendicular domain becomes homogeneous, allowing fluctuations to be represented using an eikonal (ballooning) form and treated as periodic in the perpendicular directions. Computationally, this renders the gyrokinetic field equations~(\ref{eq:Maxwell1}), (\ref{eq:Maxwell2}), and (\ref{eq:Maxwell3_LW}) diagonal in \(k_\perp\), so that gyroaveraging operations reduce to simple multiplication by Bessel functions. This greatly simplifies the numerical solution of the field equations.

\subsubsection{Global formulation of the GK equations}

In the global formulation, the same gyrokinetic and Maxwell equations as in the local (flux-tube) approach are solved, but without invoking the scale-separation assumptions that underpin the flux-tube limit. Rather than expanding equilibrium quantities about a single reference flux surface, global simulations retain the full radial dependence of the equilibrium across the computational domain. Consequently, equilibrium profiles such as \(n_{0,\sigma}(x)\), \(T_{0,\sigma}(x)\), and \(B_0(x,z)\), together with their associated gradient scale lengths, vary explicitly with radius.

This radial variation implies that the density- and temperature-gradient drives are spatially dependent, allowing both the local instability drive and the fluctuation structure to adjust self-consistently to the underlying equilibrium profiles. In addition, magnetic drift terms and magnetic shear also vary across the radial domain. From a mathematical perspective, the loss of perpendicular homogeneity means that the field equations are no longer diagonal in \(k_\perp\), precluding the use of Bessel-function representations for gyroaveraging and introducing radial coupling through the solution of the field equations.

Nonlocality arises naturally in this setting for two related reasons. First, elliptic operators such as \(-\nabla_\perp^2 \phi\) and \(\nabla_\perp^2 A_{\parallel}\) act on spatially varying coefficients associated with the equilibrium profiles. Second, gyroaveraging and pullback operators (e.g.\ \(\mathcal{K}\circ\mathcal{G}\)) sample fields at the displaced positions of particles or their guiding centres, thereby coupling fluctuations across different flux surfaces. Together, these effects break the perpendicular periodicity that underlies the flux-tube approximation.

As a result, global gyrokinetic simulations embed intrinsic nonlocality through the combined action of equilibrium variation, field solves, and gyroaveraging. This enables the capture of mesoscale phenomena such as turbulence spreading, profile shearing, and internal transport barriers, albeit at significantly greater computational cost and complexity. In particular, the loss of perpendicular periodicity necessitates the explicit treatment of radial boundary conditions, which are commonly imposed as Dirichlet conditions setting perturbations to zero at the radial boundaries. To sustain a statistically steady, gradient-driven state, artificial sources or sinks are typically introduced, as discussed in Section~\ref{subsec:boundary_conditions_numerical_sources}.

\subsection{When is global gyrokinetics needed?}

Some authors have interpreted the appearance of radially extended structures (e.g.\ streamers) in flux-tube simulations as evidence that the local approximation is breaking down and that a global treatment is therefore required~\cite{meyer2024,Li2024}. This interpretation, however, conflates the morphology of fluctuations with the ordering assumptions underlying local gyrokinetics. The local model is not invalidated by the presence of structures that span a large fraction of the perpendicular simulation domain, provided that the characteristic radial scale of the fluctuations remains small compared to the equilibrium gradient scale lengths. 

In flux-tube simulations, the radial coordinate \(x\) (normalised to the ion sound gyroradius \(\rho_s\)) and the box size \(L_x\), determined by the \(k_x\) grid and the twist-and-shift boundary condition, are numerical parameters rather than physical radial lengths. Because equilibrium quantities are frozen at a single reference surface \(x_0\), the mapping between \(x\) and real space is only meaningful locally. Consequently, structures that appear radially extended within the periodic numerical domain do not, by themselves, imply turbulence spanning an equivalent physical width, nor do they constitute evidence that global effects are dynamically important.

A genuine breakdown of the local approximation occurs only when the fundamental scale-separation assumption is violated, namely when the radial extent of the fluctuations becomes comparable to the equilibrium gradient scale lengths, such that higher-order corrections associated with finite radial variation are no longer negligible. Radial streamers observed in local simulations therefore do not, in isolation, imply the need for global gyrokinetic modelling. Rather, they may indicate proximity to a regime of strong drive or reduced stabilisation, where additional physics may become relevant. In such cases, global simulations provide a valuable tool for assessing the impact of finite profile variation, but their necessity must be established on physical grounds rather than inferred from fluctuation morphology alone. A legitimate case for global gyrokinetics must instead be grounded in experimental evidence of non-local behaviour or in theoretical arguments. 

In the absence of detailed experimental data for STEP, global simulations are valuable because they relax the scale-separation assumptions of the local model. By retaining finite-$\rho_\star$ profile variation and the resulting radial coupling in both the field equations and the drifts, they directly test whether the large fluxes predicted by local simulations~\cite{giacomin2023b,kennedy2024} persist or arise from the neglect of mesoscale physics. The goal here is precisely this comparison: to determine whether a more complete global treatment of STEP-EC-HD reproduces the large transport levels seen locally or whether global effects regulate the turbulence towards fluxes consistent with a viable operating point.

\section{The STEP-EC-HD equilibrium and numerical setup}
\label{sec:STEP-equilibrium}

\subsection{STEP-EC-HD equilibrium}

The comparison of local and global simulations motivated in Section~\ref{sec:global_GK_intro} is carried out for STEP-EC-HD, a candidate high-performance steady-state flat-top operating point designed to deliver $P_{\mathrm{fus}} \sim 1.8$~GW using electron-cyclotron heating and current drive as the only external sources of heating and current drive~\cite{tholerus2024}. The equilibrium and the kinetic profiles used as inputs here derive from JETTO integrated modelling of the STEP flat-top operational space, with predictive electron density, electron and ion temperatures and current profile, transport feedback constrained to the target normalised beta, and pellet fuelling feedback on the Greenwald fraction~\cite{tholerus2024}. Two features of the JETTO assumption-integrator approach (described in detail in \cite{tholerus2024}) to developing the reference scenario shape what follows. First, its turbulent transport was represented by simplified empirical models constrained by global confinement scalings rather than by a first-principles turbulence calculation: the confinement underpinning the design is an assumption, not a prediction. Second, the operating point is strongly self-organised: the bootstrap current fraction is $f_{\mathrm{bs}} = 0.87$, the burning plasma is heated predominantly by fusion $\alpha$ particles, and the electron-cyclotron system provides heating and current drive but no torque. The design is therefore a tightly coupled solution of pressure and current, and is highly dependent on the transport assumptions used to construct it.

Those assumptions have since been subjected to a hierarchy of increasingly complete first-principles tests. Local gyrokinetic analysis of this equilibrium identified the hybrid kinetic ballooning mode as the dominant instability and, in the absence of equilibrium flow shear, predicted turbulent fluxes at the design gradients that exceed the available sources by orders of magnitude~\cite{kennedy2023a,giacomin2023b,kennedy2024}: there is serious tension between the highest-fidelity local turbulence model and the confinement assumed in the scenario design. The global electromagnetic simulations of this paper represent the next rung of that hierarchy, testing whether global effects impact on findings from local gyrokinetics. It has not been possible to use gyrokinetic calculations to develop the STEP reference scenarios, and there is no a priori reason to expect global gyrokinetic results to agree with the reference scenario transport assumption. Global gyrokinetics relaxes the scale-separation assumed in the local model between turbulence and equilibrium lengths  (Section~\ref{subsec:local_vs_global}), so tests whether local predictions are robust to the mesoscale physics that they neglect. The remainder of this section specifies the equilibrium, resolution, and profile-control machinery with which this two-sided test is carried out; its outcome, and the implications for the operating point, are taken up in Sections~\ref{sec:linear},~\ref{sec:nonlinear}, and~\ref{sec:conclusions}.

\begin{figure}
  \centering
  \begin{subfigure}[t]{0.5\textwidth}
    \centering
    \includegraphics[]{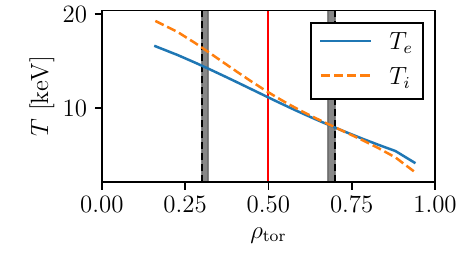}
    \subcaption{$T_{e}$ and $T_{i}$}\label{fig:profiles-a}
  \end{subfigure}\hfill
  \begin{subfigure}[t]{0.5\textwidth}
    \centering
    \includegraphics[]{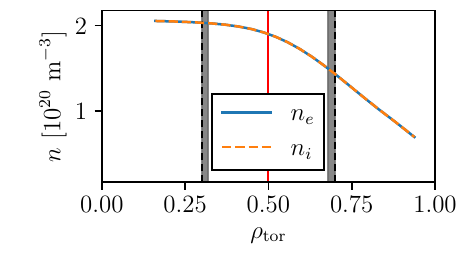}
    \subcaption{$n_{e}$ and $n_{i}$}\label{fig:profiles-b}
  \end{subfigure}

 \medskip
  \begin{subfigure}[]{0.5\textwidth}
    \centering
    \includegraphics[]{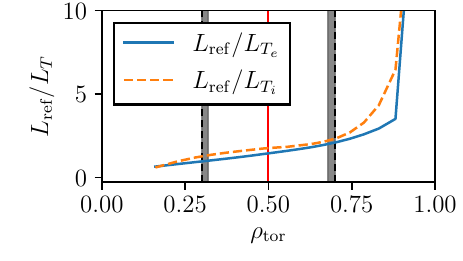}
    \subcaption{Temperature gradients}\label{fig:profiles-c}
  \end{subfigure}\hfill
  \begin{subfigure}[]{0.5\textwidth}
    \centering
    \includegraphics[]{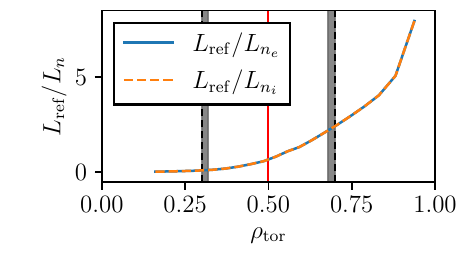}
    \subcaption{Density gradients}\label{fig:profiles-d}
  \end{subfigure}

  \medskip
  \begin{subfigure}[t]{0.5\textwidth}
    \centering
    \includegraphics[width=\textwidth]{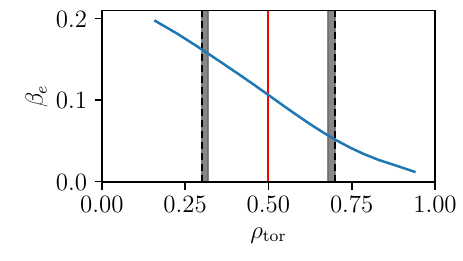}
    \subcaption{$\beta_{e}$}\label{fig:profiles-e}
  \end{subfigure}\hfill
  \begin{subfigure}[t]{0.49\textwidth}
    \centering
    \includegraphics[width=\textwidth]{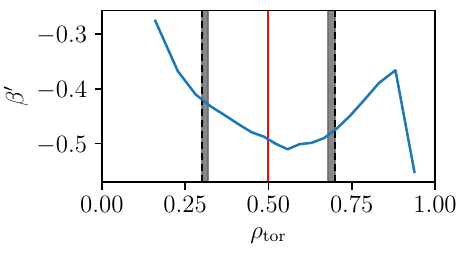}
    \subcaption{$\beta^{\prime}$}\label{fig:profiles-f}
  \end{subfigure}

  \medskip
  \begin{subfigure}[]{0.5\textwidth}
    \centering
    \includegraphics[]{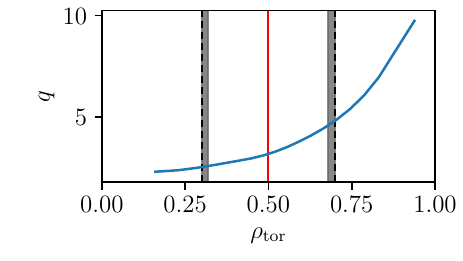}
    \subcaption{$q,$ safety factor}\label{fig:profiles-g}
  \end{subfigure}\hfill
  \begin{subfigure}[]{0.5\textwidth}
    \centering
    \includegraphics[]{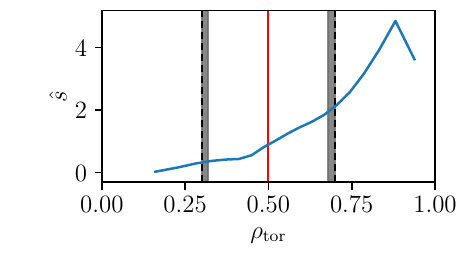}
    \subcaption{$\hat{s},$ magnetic shear}\label{fig:profiles-h}
  \end{subfigure}

  \caption{Radial profiles versus the toroidal-flux radius
  $\rho_{\mathrm{tor}}=\sqrt{\Psi_t/\Psi_{t,\mathrm{edge}}}$ for STEP-EC-HD. The red line indicates the domain centre. The black dashed lines denote the radial edges of the computational box and the grey shaded regions denote the buffer regions (see Section~\ref{subsec:boundary_conditions_numerical_sources}). Previous local studies of STEP-EC-HD~\cite{kennedy2024,kennedy2023a,Kennedy2025EM,giacomin2023b} have focused on the surface corresponding to $\rho_{\mathrm{tor}} = 0.55.$}
  \label{fig:profiles}
\end{figure}

The equilibrium profiles used in the simulations are shown in Fig.~\ref{fig:profiles} as a function of the normalised toroidal-flux radius 
$\rho_{\mathrm{tor}} = \sqrt{\Psi_t/\Psi_{t,\mathrm{edge}}}$, where $\Psi_t$ is the toroidal magnetic flux and $\Psi_{t,\mathrm{edge}}$ its value at the plasma boundary.

\begin{figure}
  \centering
  \begin{subfigure}[]{0.5\textwidth}
    \centering
    \includegraphics[]{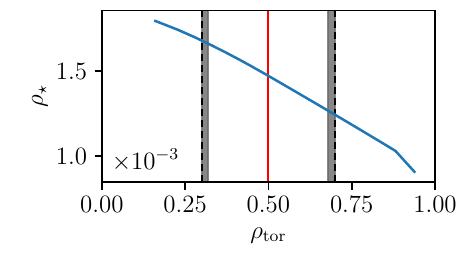}
    \subcaption{$\rho_\star,$ normalised gyroradius}\label{fig:rhostar_coll-a}
  \end{subfigure}\hfill
  \begin{subfigure}[]{0.5\textwidth}
    \centering
    \includegraphics[]{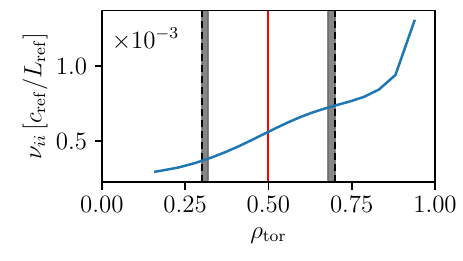}
    \subcaption{$\nu_{ii},$ ion collision frequency}\label{fig:rhostar_coll-b}
  \end{subfigure}
  \caption{Radial profiles of the normalised gyroradius $\rho_\star$ and of the ion collision frequency $\nu_{ii}$, in units of $c_{\mathrm{ref}}/L_{\mathrm{ref}}$ and with $n_i = n_e$ as in the simulations. Lines and shading as in Figure~\ref{fig:profiles}.}
  \label{fig:rhostar_coll}
\end{figure}

Table~\ref{tab:keyparams} summarises the key physical parameters of the STEP-EC-HD equilibrium and of the simulations presented in this work. The reference magnetic field $B_{\mathrm{ref}} = 2.61$~T is the toroidal field on the magnetic axis of the equilibrium, $B_{\mathrm{ref}} = F(\Psi_{\mathrm{axis}})/R_{\mathrm{axis}}$ with $F = R B_\phi$, where $R$ is the major radius and $R_{\mathrm{axis}}$ its value at the magnetic axis; since the poloidal field vanishes on axis this is the total field strength there, and it lies below the vacuum field $B_T \approx 3.2$~T because this plasma is strongly diamagnetic. The reference length $L_{\mathrm{ref}} = \sqrt{2\Psi_{t,\mathrm{edge}}/B_{\mathrm{ref}}} = 3.98$~m is an effective minor radius constructed from the toroidal flux $\Psi_{t,\mathrm{edge}}$ enclosed by the last closed flux surface (the radius of the circle that would enclose the same toroidal flux at uniform field $B_{\mathrm{ref}}$); we emphasise that it is neither the major radius nor the geometric minor radius $a = 2.01$~m. Here $\rho_\star \equiv \rho_{\mathrm{ref}}/L_{\mathrm{ref}}$ is the normalised gyroradius, where $\rho_{\mathrm{ref}} = c_{\mathrm{ref}} m_{\mathrm{ref}} / (e B_{\mathrm{ref}})$, $c_{\mathrm{ref}} = \sqrt{T_{\mathrm{ref}}/m_{\mathrm{ref}}}$, $m_{\mathrm{ref}}$ is the deuterium mass and $T_{\mathrm{ref}}$ is the electron temperature at the centre of the simulation domain ($\rho_{\mathrm{tor}} = 0.5$). The radial profile of $\rho_\star$, which varies with radius only through the electron temperature, i.e.\ $\rho_\star(\rho_{\mathrm{tor}}) = \sqrt{m_D\,T_e(\rho_{\mathrm{tor}})}/(e B_{\mathrm{ref}} L_{\mathrm{ref}})$ with $B_{\mathrm{ref}}$ and $L_{\mathrm{ref}}$ held at the fixed values above, is shown in Figure~\ref{fig:rhostar_coll}(a); it varies by less than a factor of two across the simulation domain. Figure~\ref{fig:rhostar_coll}(b) shows the corresponding profile of the ion collision frequency, which rises towards the edge as the temperature falls. The small value of $\rho_\star$ places STEP-EC-HD comfortably in the regime where local gyrokinetics would ordinarily be expected to be accurate, an observation we return to when comparing local and global fluxes in Section~\ref{sec:nonlinear}.

\begin{table}[h!]
\caption{Key parameters of the STEP-EC-HD equilibrium and of the global simulations. Equilibrium quantities are computed from the Grad--Shafranov solution; reference quantities are evaluated at the domain centre, $\rho_{\mathrm{tor}}=0.5$. $B_{\mathrm{ref}}$ is the toroidal field on the magnetic axis, and $L_{\mathrm{ref}}$ the effective minor radius constructed from the edge toroidal flux; both are defined in the main text.}
\label{tab:keyparams}
\centering
\renewcommand{\arraystretch}{1.3}
\begin{tabular}{@{}llll@{}}
\toprule
Major radius & $R_0 = 3.60$~m & Minor radius & $a = 2.01$~m \\
Aspect ratio & $A = 1.80$ & Elongation & $\kappa \approx 3.0$ \\
Plasma current & $I_p = 20.7$~MA & Toroidal field at $R_0$ & $B_T \approx 3.2$~T \\
Reference temperature & $T_{\mathrm{ref}} = 11.2$~keV & Reference density & $n_{\mathrm{ref}} = 1.9\times10^{20}$~m$^{-3}$ \\
Reference field & $B_{\mathrm{ref}} = 2.61$~T & Reference length & $L_{\mathrm{ref}} = 3.98$~m \\
$\beta_e(\rho_{\mathrm{tor}}=0.5)$ & $\beta_e \approx 0.13$ &$\rho_{\star}(\rho_{\mathrm{tor}}=0.5)$ & $\rho_\star \approx 1.5\times10^{-3}$ \\
\bottomrule
\end{tabular}
\end{table}

\subsection{Numerical setup and resolution} \label{subsec:numerical_setup}
All simulations in this study were performed with the Eulerian gyrokinetic code \texttt{GENE}, in a computational domain corresponding to the region between two flux surfaces.
In \texttt{GENE} the radial coordinate, $x$, is defined relative to the flux-surface at the centre of this domain and can be mapped to $\rho_t$ using the equilibrium metric. Unless otherwise stated, all radial positions in this work are quoted in terms of $\rho_t$.
As remarked in Section~\ref{sec:global_GK_intro}, phase space is discretised in $(x,y,z,v_\parallel,\mu)$, where $y$ is the binormal coordinate (perpendicular to both $x$ and the equilibrium magnetic field), and $z$ the coordinate along the field line. Axisymmetry in the equilibrium implies its invariance in $y$, which allows fluctuations to be treated spectrally in $y$ and  represented by binormal Fourier modes with wavenumber $k_y$ that is related to the toroidal mode number $n_0$ through
\begin{equation}
n_0 = k_y C_y,
\end{equation}
where $C_y$ is the binormal periodicity length\footnote{\parbox{\linewidth}{In global simulations the binormal coordinate $y$ is defined using a global normalisation at the low-field side, chosen such that a Fourier decomposition in $y$ corresponds directly to integer toroidal mode numbers $n_0$. This ensures consistency with the toroidal periodicity of the equilibrium and allows the turbulence to be represented as a discrete spectrum $0 < n_0 \le n_{\max}$, where the maximum resolved toroidal mode number $n_{\max}$ is set by the domain size and numerical resolution. This contrasts with local (flux-tube) simulations, where $y$ is typically normalised locally and does not correspond uniquely to a global toroidal mode number.}}.

In radially global gyrokinetic simulations, equilibrium temperature and density profiles generally vary with radius, so that the local Maxwellian distribution,
\begin{equation}
f_{M,\sigma}(x,v)\propto \exp\!\left[-\frac{v^2}{v_{\mathrm{th},\sigma}^2(x)}\right], 
\qquad 
v_{\mathrm{th},\sigma}(x)=\sqrt{\frac{2T_\sigma(x)}{m_\sigma}},
\end{equation}
changes significantly across the simulation domain. As a result, a single, radially uniform velocity-space grid must be sized to accommodate the largest thermal velocity in the system, leading to severe over-resolution of $v_\parallel$ and $\mu$ in regions where $T_\sigma(x)$ is smaller. To avoid this inefficiency, \texttt{GENE} can employ block-structured grids (BSG) \cite{jarema2015}, in which the radial domain is divided into a finite number of radial blocks, each associated with a velocity-space grid scaled to a representative local thermal velocity within that block. All radial grid points inside a given block share the same $(v_\parallel,\mu)$ discretisation, chosen to span several local thermal widths of $f_{M,\sigma}(x,v)$ with sufficient resolution. Adjacent radial blocks overlap over a finite radial extent, allowing moments and electromagnetic fields to be smoothly coupled between blocks and preventing artificial discontinuities at block interfaces. This approach significantly reduces the total number of velocity-space grid points required, while retaining accuracy in the presence of strong radial profile variation, making radially global gyrokinetic simulations computationally tractable.

Resolution studies were carried out in configuration space, where convergence was tested by enlarging the simulation domain in $x$ and by varying the number of radial, binormal, and field-line grid points $(n_x,n_y,n_z)$. These tests confirmed that linear growth rates and eigenmode structures were well resolved and emphasised the need to capture low-$k_y$ modes, which dominate transport in STEP-like regimes, while maintaining adequate high-$k_y$ resolution to resolve small scale physics. The final nonlinear runs used the minimal resolution that reproduced converged fluxes within approximately $10\%$, balancing accuracy with computational cost. A repeat of the nominal nonlinear case on a finer grid reproduced the same behaviour. The nominal simulation parameters are summarised in Table~\ref{tab:resolution}, with all symbols defined in the caption.

\begin{table}[h!]
\caption{Nominal resolution parameters for linear (L) and nonlinear (NL) simulations in local (LOC) and global (G) geometry. Here $n_{k_x}$ ($n_x$) is the spectral (real) radial resolution, $n_{k_y}$ the number of binormal Fourier modes, $n_z$ the points along the field line, and $n_{v_\parallel}$ and $n_\mu$ the velocity-space resolutions. $L_x$ and $L_y$ are the local domain sizes in gyroradii, $L_{x}/a$ and $x/a$ the width and centre of the global domain in minor radii, and $n_0^{\min}$ and $n_0^{\max}$ the smallest non-zonal and largest toroidal mode numbers evolved. For the NL (LOC) simulations the twist-and-shift parallel boundary condition quantises the radial box size to integer multiples of $1/(\hat{s}\,k_y^{\mathrm{min}})$, and $L_x$ is chosen independently on each flux surface as the multiple closest to the target size; the values shown are those of the $\rho_{\mathrm{tor}}=0.50$ surface.}
\label{tab:resolution}
\centering
\renewcommand{\arraystretch}{1.5}
\begin{tabular}{@{}lcccccccccc@{}}
\toprule
\textbf{Simulation} & \( n_{k_x} \) & \(n_{k_y} \) & \( n_{z} \) & \( n_{v_\parallel} \) & \( n_{\mu} \) & \( L_{x} \) & \( k_{x}^{\mathrm{min}} \) & \( L_y \) & \( n_{0}^{\mathrm{min}} \) & \( n_{0}^{\mathrm{max}} \) \\ \midrule
L (LOC)          & 5          & 1           & 64                    & 32    & 16           & -             & -      & - & - & -                 \\
NL (LOC)          & 128          & 32             & 64                    & 28    & 14           & 534            & 0.01                    & 338  & 2 & 128  \\ \\ \textbf{Simulation} & \( n_{x} \) & \(n_{k_y} \) & \( n_{z} \) & \( n_{v_\parallel} \) & \( n_{\mu} \) &  \(x/a \) & \( L_{x}/a \) & \( n_{0}^{\mathrm{min}} \) & \( n_{0}^{\mathrm{max}} \) \\ \midrule
L (G)         & 1024          & 1             & 280                    & 56    & 28           & 0.5           & 0.4                    & -  & -  \\
NL (G), production         & 256          & 64             & 70                    & 48   & 24         & 0.5          & 0.4                    & 4 & 252  \\
NL (G), $\beta_e$ comparison         & 128          & 64             & 70                    & 48   & 24           & 0.5          & 0.4                    & 2 & 126  \\
NL (G), fine-grid check         & 512          &64             & 70                    &64   & 32           & 0.5          & 0.4                    & 4 & 256  \\
 \bottomrule
\end{tabular}
\end{table}

Nominally, STEP integrated modelling includes six species (electrons, deuterium, tritium, helium ash, a heavy impurity, and fast $\alpha$ particles). In the present work, unless explicitly stated otherwise, only two kinetic species are retained (electrons and deuterium), with $\alpha$ particles neglected. This simplification is made for computational tractability, since each additional kinetic species adds a further velocity-space grid to an already expensive global electromagnetic simulation, and for continuity with the earlier local studies of this equilibrium~\cite{kennedy2023a,giacomin2023b,kennedy2024}, which used the same two-species treatment and against which we compare directly. The physical consequence is that dilution of the main-ion pressure-gradient drive by impurities, helium ash, and fast $\alpha$ particles, which is present in the integrated modelling, is neglected here; this dilution is expected to be stabilising for the pressure-gradient-driven modes studied in this work, and its effect on this operating point is examined in forthcoming work~\cite{bokshi2026}. The fuel is represented by a single ion species throughout: a pure deuteron, with $Z = 1$ and $n_i = n_e$. The electron-ion mass ratio is set to the physical deuterium value, $m_i/m_e \approx 3672$. Electromagnetic fluctuations are retained through the electrostatic potential $\phi_1$, the parallel vector potential $A_{1,\parallel}$, and the parallel magnetic field perturbation $B_{1,\parallel}$. The latter is included via the long-wavelength approximation introduced in \cite{sheffield2024}, following the demonstration in \cite{kennedy2024} that $B_{1,\parallel}$ is essential for capturing the dominant long-wavelength electromagnetic instabilities, labelled hKBMs, in STEP-like regimes.

Unless stated otherwise, simulations are carried out without equilibrium $E \times B$ flow shear, so that the only shearing of turbulence arises from zonal flows generated by the turbulence. Since STEP-EC-HD is heated by electron-cyclotron waves alone, there is no NBI torque, and the expected equilibrium radial electric field follows from the radial force balance of the main-ion species,
\begin{equation}
E_r \;=\; \frac{1}{Z_i e n_i}\,\frac{\mathrm{d}p_i}{\mathrm{d}r} \;+\; v_{\varphi,i}B_\theta \;-\; v_{\theta,i}B_\varphi,
\label{eq:force_balance}
\end{equation}
where $Z_i e$ is the ion charge, $n_i$ and $p_i$ are the ion density and pressure, $r$ is the minor radius, $v_{\varphi,i}$ and $v_{\theta,i}$ are the toroidal and poloidal ion flow velocities, and $B_\varphi$ and $B_\theta$ are the toroidal and poloidal components of the magnetic field. The first (diamagnetic) term survives even in the absence of bulk rotation. Evaluating this term at the outboard midplane across the simulation domain gives $E_r \approx -12$ to $-28$~kV/m. For input to \texttt{GENE}, this field is expressed as the equivalent toroidal angular frequency, with the subscript omp denoting evaluation at the outboard midplane,
\begin{equation}
\Omega_{\mathrm{tor}}(x) \;=\; \left.\frac{E_r}{R\,B_\theta}\right|_{\mathrm{omp}},
\label{eq:omega_tor}
\end{equation}
that is, the rigid toroidal rotation that would produce the same radial field through the $v_{\varphi}B_\theta$ term of (\ref{eq:force_balance}), with the major radius $R$ and $B_\theta$ evaluated at the outboard midplane. The associated $E\times B$ shearing rate is the Waltz-Miller rate~\cite{waltzmiller1999}
\begin{equation}
\gamma_E \;=\; \left|\frac{x}{q}\,\frac{\mathrm{d}\Omega_{\mathrm{tor}}}{\mathrm{d}x}\right|\frac{L_{\mathrm{ref}}}{c_{\mathrm{ref}}},
\label{eq:gammaE}
\end{equation}
which for the diamagnetic field gives $\gamma_E \approx 10^{-3}$ to $5\times10^{-3}\,c_{\mathrm{ref}}/L_{\mathrm{ref}}$, roughly two orders of magnitude below the peak linear growth rates ($\gamma_{\mathrm{max}} \approx 0.36\,c_{\mathrm{ref}}/L_{\mathrm{ref}}$ at the binormal scales considered in this work, Section~\ref{sec:linear}). The smallness of this floor is itself a physically important property of the operating point. To bracket the possible role of equilibrium flow shear, the nonlinear simulations of Section~\ref{sec:nonlinear} impose rotation profiles at exactly this diamagnetic level and, separately, a synthetic profile with radially uniform $\gamma_E = \gamma_{\mathrm{max}}$, the level at which the Waltz quench rule predicts full suppression of the turbulence~\cite{kinsey2005}. A published local flux-tube scan in $\gamma_E$ at the same operating point~\cite{giacomin2023b} brackets the shearing rate at which the turbulence is actually quenched. The impact of these equilibrium shear levels on the nonlinear global simulations is addressed in Section~\ref{sec:nonlinear}.

\subsection{Boundary conditions and numerical sources} \label{subsec:boundary_conditions_numerical_sources}

In contrast to local (flux-tube) simulations, where periodic boundary conditions enforce fixed background gradients by construction, radially global simulations require explicit control of both boundary behaviour and the evolution of the flux-surface-averaged profiles. In \texttt{GENE}, Dirichlet boundary conditions are applied at the radial edges in $x$, such that the perturbed distribution function $\delta f_\sigma$ vanishes there. This condition eliminates all fluctuating moments at the boundaries and, through the field equations, enforces vanishing electrostatic and electromagnetic perturbations at the domain edges. While these boundary conditions fix the profiles locally at the radial boundaries, they should not constrain the interior evolution of the profiles.

To ensure that the imposed boundary conditions do not contaminate the turbulent dynamics in the interior, buffer regions are introduced near the radial edges. Within these regions, outgoing fluctuations are absorbed using Krook-type damping operators acting on the perturbed distribution function $g_{1\sigma}$,
\begin{equation}
\frac{\partial g_{1\sigma}}{\partial t}
= \hat{V}(g_{1\sigma}) - \nu_{\mathrm{Krook}}(x)\, g_{1\sigma},
\end{equation}
where $\hat{V}$ denotes the standard gyrokinetic operator and $\nu_{\mathrm{Krook}}(x)$ is a spatially dependent damping rate. The damping is chosen to vanish in the central region of the domain and to increase smoothly towards the radial boundaries, typically over the outermost $\sim 5\%$ of the radial extent. Its maximum value is taken to be comparable to the largest linear growth rate in the system, ensuring efficient absorption of fluctuations without affecting the core turbulence. The buffer regions therefore act to stabilise the numerical solution and to isolate the physical turbulent region from the artificial boundaries.

Even with fixed boundary conditions and buffer damping, without internal sources turbulent transport would progressively relax the internally prescribed profiles towards marginality. To sustain a statistically steady state, numerical source terms act only on the flux-surface-averaged component of the distribution function, maintaining the average profiles close to their targets while leaving the fluctuating dynamics of $\delta \! f_\sigma$ unaffected; the scheme is a two-term feedback controller and is detailed in~\ref{app:sources}. In our simulations the Krook relaxation rates are held fixed at $c_{k,\mathrm{heat}} = c_{k,\mathrm{part}} = 0.01\,c_{\mathrm{ref}}/L_{\mathrm{ref}}$, roughly $3\%$ of the peak linear growth rate ($\gamma_{\max} \approx 0.36\,c_{\mathrm{ref}}/L_{\mathrm{ref}}$ at $n_0 \approx 100$, see Figure~\ref{fig:1d_local_global_growthrates}): the feedback is slow compared with the turbulence but fast compared with the mean profile evolution. The resulting instantaneous profile deviations, the sensitivity to $c_k$, and the source-free limit $c_k = 0$ are discussed in~\ref{app:sources}.

We begin by comparing local and global linear simulations in order to assess whether global effects play a significant role in STEP-EC-HD and to validate the accuracy of the long-wavelength solver for capturing the behaviour of the modes with smallest toroidal mode numbers, $n_{0}$, i.e., the modes with the longest wavelength.  

\section{Global linear simulations of electromagnetic instabilities in STEP-EC-HD} \label{sec:linear}

\subsection{Comparing local and global gyrokinetic simulations}
\label{subsec:comparing_local_global_GK_simulations}

Global simulations of the plasma core are challenging for many reasons. One reason is that the kinetic profiles tend to be flatter in the core and steeper towards the edge (see Figure~\ref{fig:profiles}). As a result, the gyrokinetic drive terms are (typically) monotonically increasing functions of radius. As can be seen from  Figure~\ref{fig:profiles}, for STEP-EC-HD both the temperature and density gradients grow steadily towards the outer regions of the global simulation domain (the right-hand-side of the global box). In the simulation the gradients are reduced to zero in the buffer zones by Krook operators (grey shaded regions). Practically, this means that an eigenvalue obtained from a local calculation for a particular mode number, say $n_{0}=n_{1},$ need not agree with a global calculation centred at the same radial position, since the drive of the mode in the global calculation instead represents an average over different radii where the drive strength varies. For this reason, a direct comparison between global and local eigenvalues (and eigenmodes) requires a clear and physically motivated choice of the radial position at which the local calculation is performed. A natural choice is the radial location at which the global eigenfunction attains its maximum amplitude. At this position, the mode experiences the strongest effective drive and therefore contributes most significantly to the global instability. 

\begin{figure}
  \centering
  \begin{minipage}[t]{0.5\textwidth}
    \centering
    \includegraphics[]{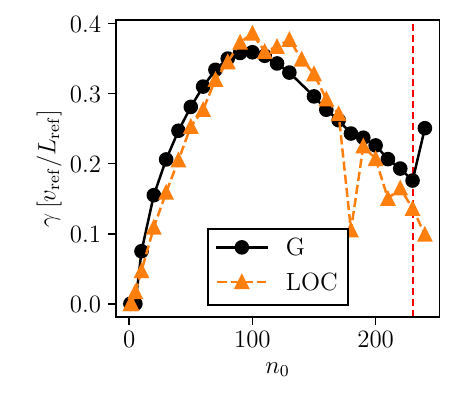}
  \end{minipage}\hfill
  \begin{minipage}[t]{0.5\textwidth}
    \centering
    \includegraphics[]{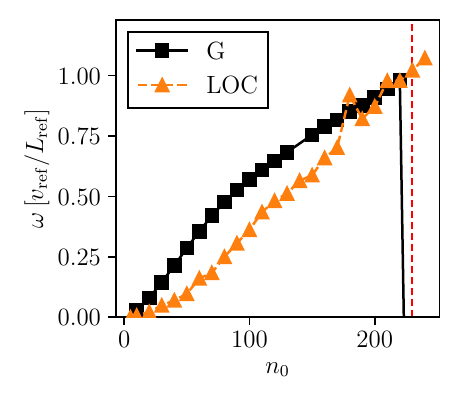}
\end{minipage}
  \caption{Growth rate (left-hand side) and mode frequency (right-hand side) against toroidal mode number $n_0$ from \texttt{GENE} linear simulations of STEP-EC-HD, for global (G) and local (LOC) calculations. Each local calculation is at the radius where the global eigenmode peaks, and the red vertical line marks the mode number above which that eigenmode's radial width vanishes (Figure~\ref{fig:1d_global_localisation}). With the binormal normalisation used here, $n_0 = 100$ corresponds to $k_\theta\rho_{\mathrm{ref}} \approx 0.93$. Positive real frequencies correspond to rotation in the ion diamagnetic direction.}

  \label{fig:1d_local_global_growthrates}
\end{figure}

\begin{figure}
  \centering
  \begin{minipage}[t]{0.5\textwidth}
    \centering
    \includegraphics[]{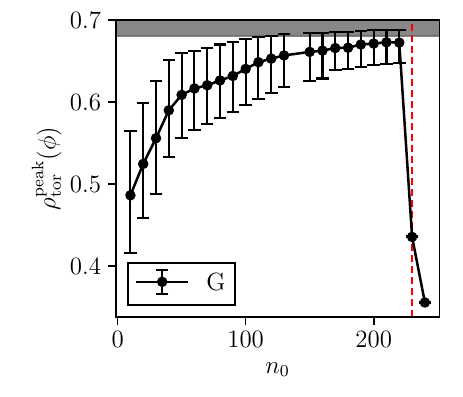}
  \end{minipage}\hfill
  \begin{minipage}[t]{0.5\textwidth}
    \centering
    \includegraphics[]{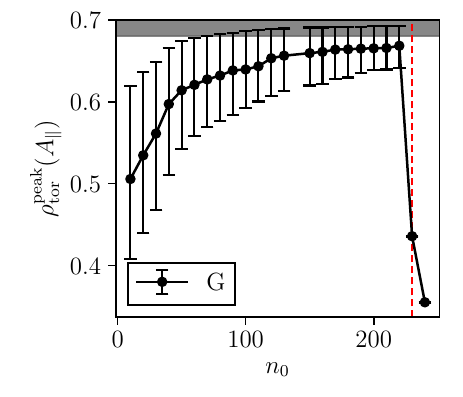}
\end{minipage} \\
  \begin{minipage}[t]{0.5\textwidth}
    \centering
    \includegraphics[]{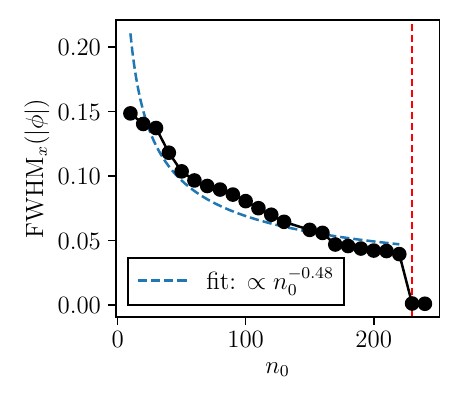}
  \end{minipage}\hfill
  \begin{minipage}[t]{0.5\textwidth}
    \centering
    \includegraphics[]{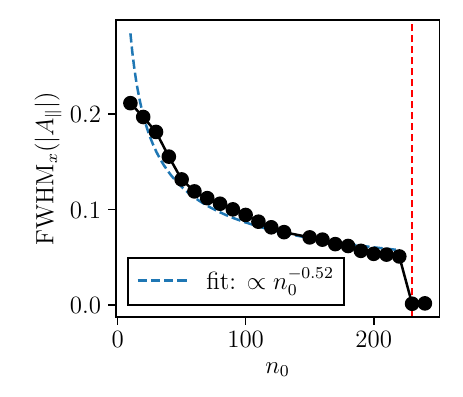}
\end{minipage}
  \caption{Radial localisation properties of unstable global modes as a function of toroidal mode number $n_0$. (top) Radial position $\rho_{\mathrm{tor}}^{\mathrm{peak}}$ at which the electrostatic potential (left-hand side) and parallel vector potential (right-hand side) attain their maximum amplitudes, $\rho_{\mathrm{tor}}^{\rm{peak}}(\phi) = x(\arg\max_{x,z}|\phi|)$  and $\rho_{\mathrm{tor}}^{\rm{peak}}(A_\parallel) = x(\arg\max_{x,z}|A_\parallel|).$ Whiskers denote the radial extent of the mode (i.e., the FWHM). The buffer regions are shaded in grey. (bottom) Radial full width at half maximum, $\mathrm{FWHM}_x(\phi)$, defined from the envelope
$\max_z|\phi(x,z)|$ (left-hand side) and $\mathrm{FWHM}_x(A_\parallel)$, defined from the envelope
$\max_z|A_\parallel(x,z)|$. Power law fits are shown as dashed blue lines. Only unstable modes are shown. The red vertical line marks a cut-off for the maximum toroidal mode number that can be resolved where the FWHM becomes zero.}
  \label{fig:1d_global_localisation}
\end{figure}

Figure~\ref{fig:1d_local_global_growthrates} shows the growth rate (left-hand side) and real frequency (right-hand side) as functions of the toroidal mode number obtained from linear \texttt{GENE} simulations, comparing global and local modes of operation. The global simulation domain spans \(0.3 \leq \rho_{\mathrm{tor}} \leq 0.7\). For each toroidal mode number, the corresponding local calculation is performed at the radial position \(\rho_{\mathrm{tor}}^{\mathrm{peak}}\), defined as the location at which the global electrostatic potential attains its maximum amplitude [see Figure~\ref{fig:1d_global_localisation} (top, left-hand side)]. The real frequencies are positive across the unstable range, corresponding to rotation in the ion diamagnetic direction, as expected for modes in the KBM family and consistent with the hKBM identified in local studies of this equilibrium~\cite{kennedy2023a}. Over a broad range of toroidal mode numbers, the local simulations reproduce both the magnitude and spectral trends of the global growth rates and mode frequencies with good quantitative agreement. This indicates that, for these modes, the dominant contribution to the global instability arises from physics local to the region where the mode amplitude is maximal, and that non-local effects associated with the finite radial extent of the global eigenfunction are subdominant. Noticeable deviations appear only at higher toroidal mode numbers, where the global modes become increasingly localised towards the outer part of the simulation domain (where the drive is strongest) and can perhaps be more strongly influenced by proximity to the buffer regions [see Figure~\ref{fig:1d_global_localisation} (top)]. We note that a more comprehensive (though computationally more expensive) approach for selecting a suitable local simulation to compare with is to compare with the maximum local result over the full radial domain of the global simulation, rather than at a single radial location. With this definition, Figure~\ref{fig:1d_local_global_growthrates} would not show global results exceeding local ones and would display improved agreement in frequency.

The radial extent of the global eigenmodes found in STEP-EC-HD can span a substantial fraction of the core. To further characterise the radial localisation of global eigenmodes, we define a radial full width at half
maximum (FWHM) of the electrostatic potential $\phi(\rho_{\mathrm{tor}},z)$ (and analogously for the parallel vector potential). The two-dimensional field is first reduced to a one-dimensional radial envelope by taking the maximum
over the field-aligned coordinate,
\begin{equation}
\Phi(\rho_{\mathrm{tor}}) \equiv \max_{z} |\phi(\rho_{\mathrm{tor}},z)|.
\end{equation}
The radial FWHM is then defined as
\begin{equation}
\mathrm{FWHM}(\phi)
\equiv
\rho_{\mathrm{tor},+} - \rho_{\mathrm{tor},-},
\qquad
\Phi(\rho_{\mathrm{tor},\pm}) \equiv \frac{1}{2}\max_{\rho_{\mathrm{tor}}} \Phi(\rho_{\mathrm{tor}}),
\end{equation}
where $\rho_{\mathrm{tor},\pm}$ denote the outermost and innermost radial locations at which the envelope exceeds half of its maximum value. $\mathrm{FWHM}(\phi)$ provides a robust measure of radial mode localisation that is insensitive to fine-scale oscillatory structure along the field line. Modes for which the FWHM spans many radial grid points are well resolved and only weakly influenced by finite-domain effects. Conversely, modes whose FWHM is comparable to the radial grid spacing or to the numerical domain size may be under-resolved or artificially constrained by boundary conditions. In this sense, the radial FWHM serves as a practical diagnostic for distinguishing physically meaningful global modes from structures whose apparent localisation is dominated by numerical resolution. 

Figure~\ref{fig:1d_global_localisation} (bottom) shows the FWHM as a function of toroidal mode number. At low toroidal mode number, the radial full width at half maximum (FWHM) is relatively large, indicating that these modes are intrinsically global, with radial structures that extend over a substantial fraction of the simulated domain and are sensitive to equilibrium variation across neighbouring flux surfaces. As the toroidal mode number \(n_0\) is increased, the FWHM decreases, reflecting enhanced radial localisation and a progressive transition towards more localised mode structure (as well as beginning to encroach on the buffer regions). The measured $\mathrm{FWHM}_x$ scalings [power law fits are shown in Figure~\ref{fig:1d_global_localisation} (bottom)] are consistent with the theoretical prediction of \cite{Dickinson2014}, which for fixed equilibrium parameters implies that the mode's radial width scales as
\begin{equation}
\Delta_x \;\propto\; \frac{1}{\sqrt{n\,q'}}\,,
\end{equation}
where $q'$ is the radial derivative of the safety factor. This scaling arises from the rational grid spacing scaling as $1/(nq^{\prime})$ and the number of rational surfaces that are coupled increasing as $\sqrt{nq^\prime}.$ This scaling facilitates a quantitative prediction of the global mode's radial extent. Eventually, the mode becomes sufficiently radially localised that it is no longer adequately resolved on the numerical grid, which represents an inherent limitation of global simulations. In the present cases, this does not constitute the limiting factor; instead, the dominant constraint arises from the spacing between neighbouring rational surfaces. This nevertheless provides the first indication that global simulations are only able to resolve fluctuations down to a minimum radial wavelength, as discussed in Section~\ref{subsubsec:global_constraints}.

\subsection{Linear simulations of long-wavelength instabilities}

Long-wavelength electromagnetic instabilities are expected to dominate turbulent transport in STEP-like plasmas~\cite{kennedy2024,giacomin2023b}. Figure~\ref{fig:1d_local_global_growthrates} shows excellent agreement between growth rate and frequency spectra from both local initial value calculations and global initial value calculations using the long-wavelength $\dbp$ solver of~\cite{sheffield2024}, suggesting that global \texttt{GENE} is able to capture the hKBMs identified in~\cite{kennedy2023a}. Here we verify that these modes are indeed the global analogue of the hKBM by examining the eigenstructures of these modes.  

\begin{figure}
    \centering
    \includegraphics[width=\textwidth]{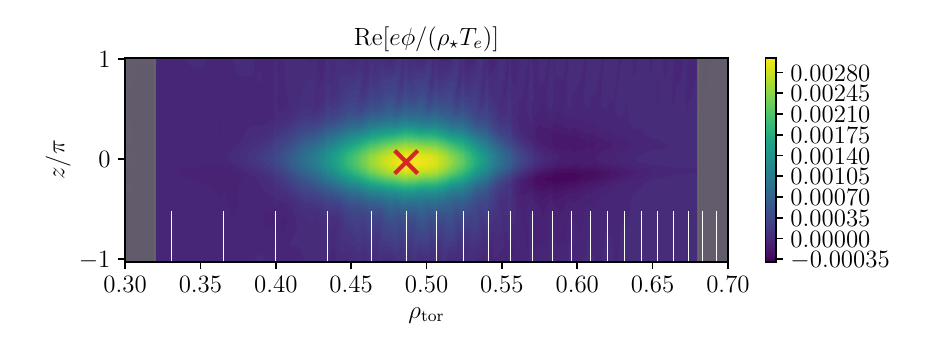}
    \includegraphics[width=\textwidth]{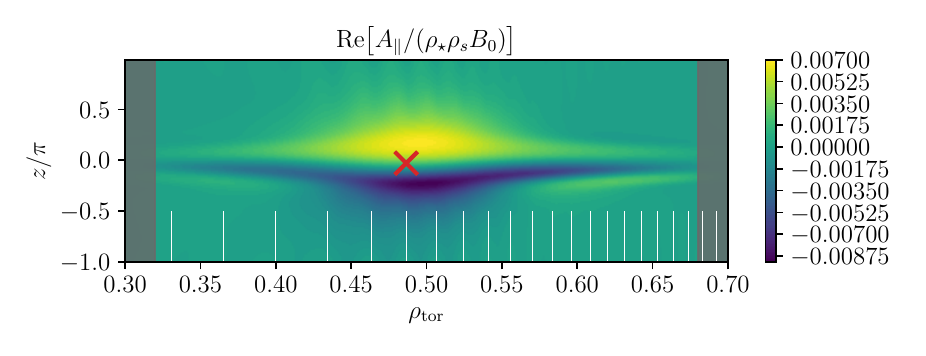}
    \includegraphics[width=\textwidth]{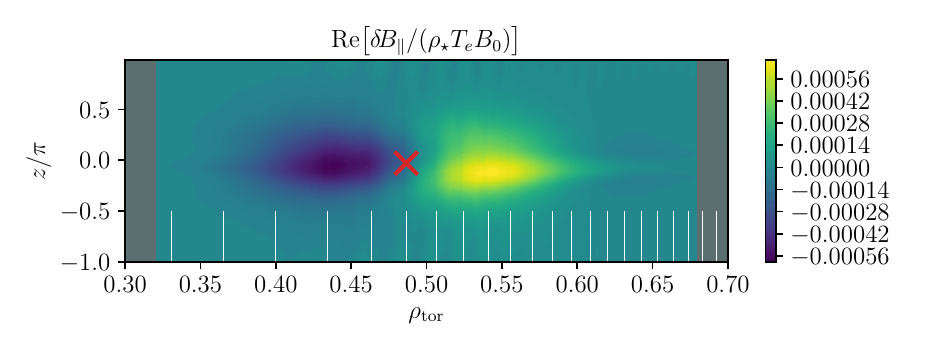}
  \caption{[Long-wavelength solver for $\dbp$] Real part of the fluctuating electrostatic potential (top), parallel magnetic vector potential (middle), and parallel magnetic perturbation (bottom) as functions of the radial coordinate $\rho_{\text{tor}}$ and the field-line-following coordinate $z$ at the final time step of a linear calculation. A global phase factor is applied to all fields such that the complex phase of the maximum electrostatic potential is removed. Results are shown for a mode with toroidal mode-number $n_0=10.$ White lines mark the location of rational surfaces where $q = m/10, \, m \in \mathbb{Z}^{+}$. The buffer regions are shaded grey. The mode peaks in the centre of the computational box (red cross) and has ballooning parity.}
  \label{fig:2d_modestructures_n10}
\end{figure}

\begin{figure}
  \centering
\includegraphics[]{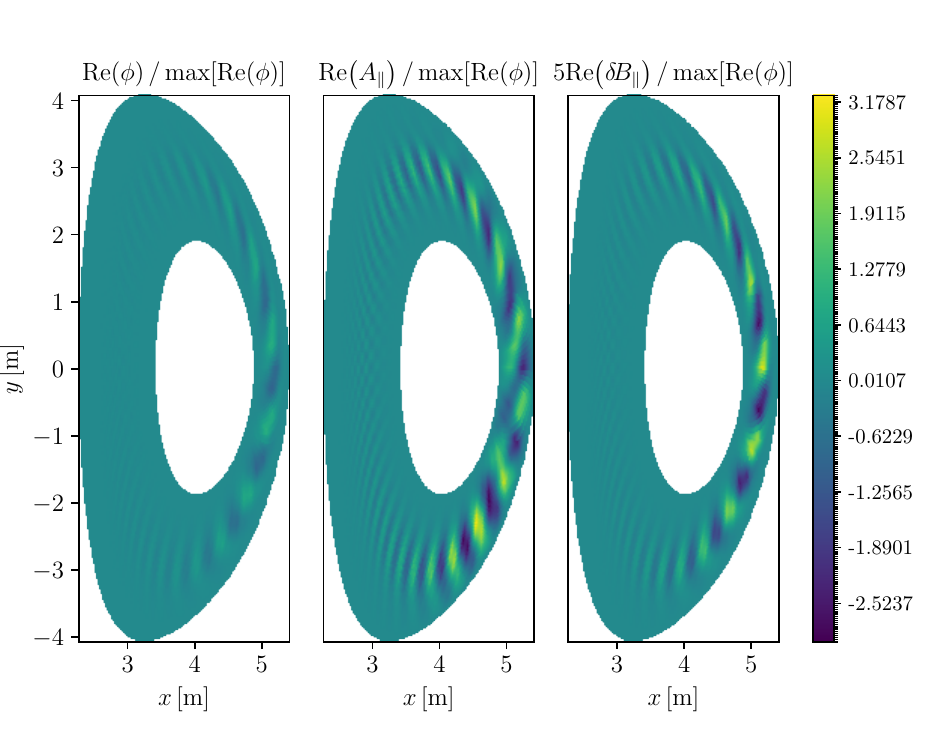}
  \caption{[Long-wavelength solver for $\dbp$] Poloidal cross sections of the real part of the fluctuating electrostatic potential (left-hand side), parallel magnetic vector potential (middle), and parallel magnetic perturbation (right-hand side) for $n_0 = 10$. Note that
the aspect ratio of the panels corresponds to that of the STEP-EC-HD cross section.}
  \label{fig:modestuctures_n10_poloidal}
\end{figure}

Figure~\ref{fig:2d_modestructures_n10} shows the real part of the fluctuating electrostatic potential, parallel magnetic vector potential, and parallel magnetic field for the $n_0=10$ mode, plotted as functions of $\rho_{\text{tor}}$ and the field-line-following coordinate $z$ (the corresponding poloidal cross sections appear in Figure~\ref{fig:modestuctures_n10_poloidal}). The eigenfunctions peak in the centre of the computational domain (indicated by the red cross), away from the buffer regions where the drive is suppressed, demonstrating that the structure is physical and well-resolved. The mode exhibits some radial localisation around the centre of the domain and is narrow in $z$ (ballooning). 

\begin{figure}
  \centering
  \begin{minipage}[t]{0.33\textwidth}
    \centering
    \includegraphics[]{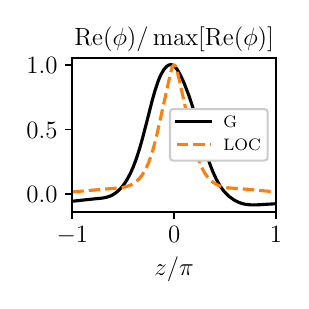}
  \end{minipage}\hfill  \begin{minipage}[t]{0.33\textwidth}
    \centering
    \includegraphics[]{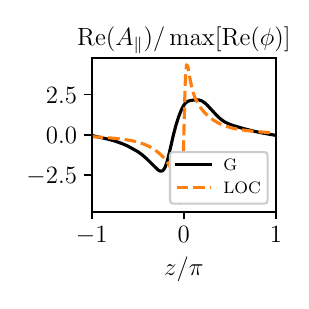} \end{minipage} \hfill
  \begin{minipage}[t]{0.33\textwidth}
    \centering
    \includegraphics[]{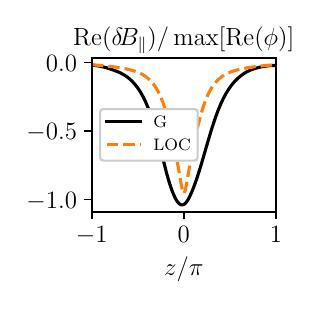}
\end{minipage}
  \caption{[Long-wavelength solver for $\dbp$] Eigenfunctions of the real part of the fluctuating electrostatic potential (left-hand side), parallel magnetic vector potential (middle), and parallel magnetic perturbation (right-hand side) for $n_0 = 10$. Eigenfunctions are shown for both the global code (G) and from the local code (LOC). The global eigenfunction is sampled along the ballooning field line passing through the radial position of maximum $|\phi|$, and the local eigenfunction is evaluated at the corresponding flux-surface location so that both represent the same physical radius. 
}
  \label{fig:1d_modestuctures_n10}
\end{figure}

The corresponding one-dimensional eigenfunctions are shown in Figure~\ref{fig:1d_modestuctures_n10}. Specifically, these are slices of \(\phi(\rho_{\mathrm{tor}},z)\), \(A_\parallel(\rho_{\mathrm{tor}},z)\), and \(\delta\!B_\parallel(\rho_{\mathrm{tor}},z)\) taken at the radial location \(\rho_{\mathrm{tor}}^{\mathrm{peak}}\) corresponding to the peak of the global mode amplitude. These global eigenfunctions are directly compared\footnote[1]{Wherever global and local eigenfunctions are compared they are first brought into a common gauge following the procedure outlined in~\ref{app:eigenfunction_comparison}} with the corresponding local flux-tube eigenfunctions evaluated at the same radial location. In both cases, all fields display the expected parity properties of the hKBM previously identified under STEP-relevant conditions~\cite{kennedy2023a}. The local eigenfunctions are visibly more localised along the field line than their global counterparts. The global eigenmode is a superposition of coupled poloidal harmonics centred on neighbouring rational surfaces, and its radial envelope samples a finite range of gradients and safety-factor values, so the parallel structure represents an average over this envelope, which broadens it relative to a local calculation performed at the single most unstable radius. This difference is expected to vanish in the limit $\rho_\star \to 0$. Notwithstanding this difference, the close agreement between the local and global mode eigenmodes and eigenfunctions at \(n_0=10\) provides confidence that the long-wavelength global solver is accurately resolving the same dominant instability.

\subsection{On the need for parallel perturbation of the magnetic field}

\begin{figure}
  \centering
  \begin{minipage}[t]{0.5\textwidth}
    \includegraphics[]{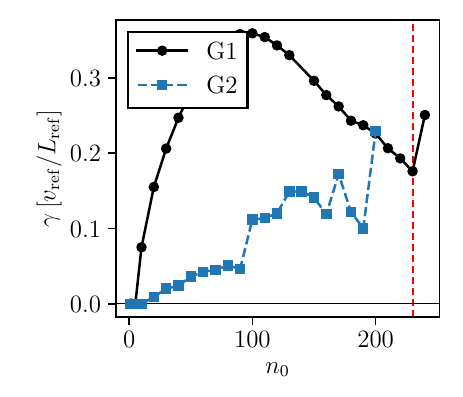}
  \end{minipage}\hfill
  \begin{minipage}[t]{0.5\textwidth}
    \centering
    \includegraphics[]{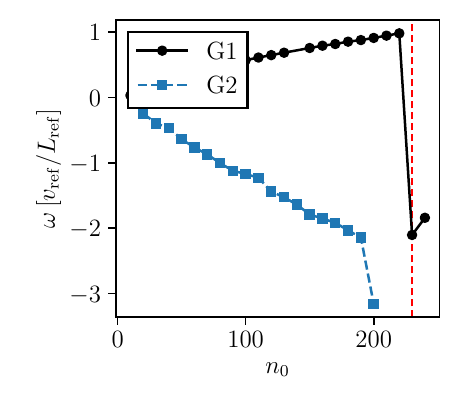}
\end{minipage}
  \caption{Growth rate (left-hand side) and mode frequency (right-hand side) as functions of the toroidal mode number $n_0$ from \texttt{GENE} linear simulations of STEP-EC-HD. Results are shown for global calculations that include (G1) or exclude (G2) $\dbp$ fluctuations.}

  \label{fig:1d_local_global_growthrates_bparcomp}
\end{figure}

It was demonstrated in~\cite{kennedy2023a} that parallel magnetic perturbations, $\dbp,$ are essential for correctly capturing the long-wavelength behaviour of the hKBM in local gyrokinetic simulations. This is also true in global simulations. Figure~\ref{fig:1d_local_global_growthrates_bparcomp} shows the growth rate and mode frequencies as functions of the toroidal mode number from global calculations with (G1 - using the long-wavelength model) and without (G2) $\dbp$ (note that these simulations have the same resolution as those with $\dbp$ and that the full drift velocity is retained in all cases, i.e., no MHD approximation is used~\cite{kennedy2024}). Without $\dbp$ fluctuations, the global code is not able to correctly identify the hKBM, instead returning modes that are much more weakly growing and propagate in the electron diamagnetic direction. This behaviour is consistent with previous local gyrokinetic results, in which neglecting \(\delta B_\parallel\) stabilises the hKBM and causes the initial-value solver to converge instead to a subdominant microtearing mode (MTM)~\cite{kennedy2023a}. 

\begin{figure}
    \centering
    \includegraphics[]{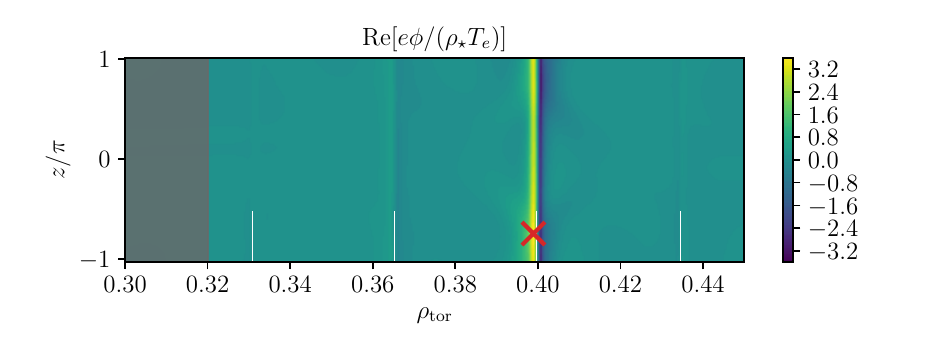}
    \includegraphics[]{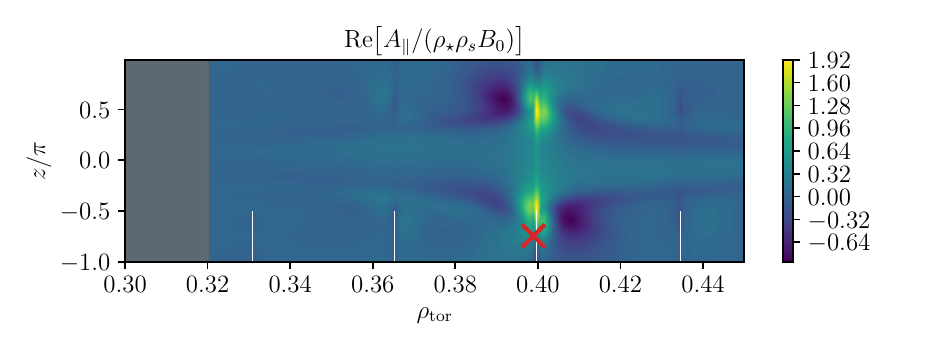}
  \caption{[No evolution of $\dbp$] Real part of the fluctuating electrostatic potential (top) and parallel magnetic vector potential (bottom) as functions of the radial coordinate $\rho_{\text{tor}}$ and the field-line-following coordinate $z$ at the final time step of a linear calculation. A global phase factor is applied to all fields such that the complex phase of the maximum electrostatic potential is removed. Results are shown for a mode with toroidal mode-number $n=10.$ White lines mark the location of rational surfaces where $q = m/10, \, m \in \mathbb{Z}^{+}$. The mode is very strongly localised to a single rational surface (red cross) [note that the figure is zoomed in on the rational surface where the mode peaks].
}
  \label{fig:2d_modestructures_n10_noBpar}
\end{figure}

Based on this correspondence, we infer that the modes obtained in the global calculation without \(\delta \! B_\parallel\) are likely MTMs rather than hKBMs. In the following, we investigate the nature of these modes in more detail to confirm this interpretation. Figure~\ref{fig:2d_modestructures_n10_noBpar} shows the absolute value of the fluctuating electrostatic potential and parallel magnetic vector potential for the $n_0=10$ mode from the simulation where $\dbp$ is neglected, plotted as functions of $\rho_{\text{tor}}$ and the field-line-following coordinate $z.$ Unlike the case using the long-wavelength $\dbp$ solver (Figure~\ref{fig:2d_modestructures_n10}) the mode is strongly localised to a single rational surface. 

\begin{figure}
    \centering
    \includegraphics[width=\textwidth]{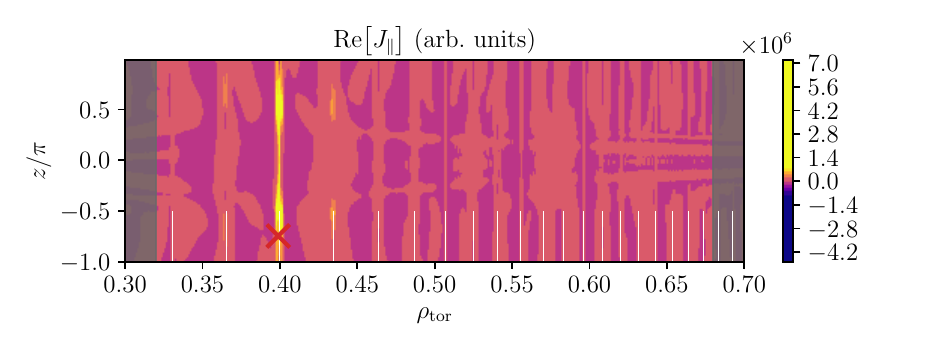}
    \includegraphics[width=\textwidth]{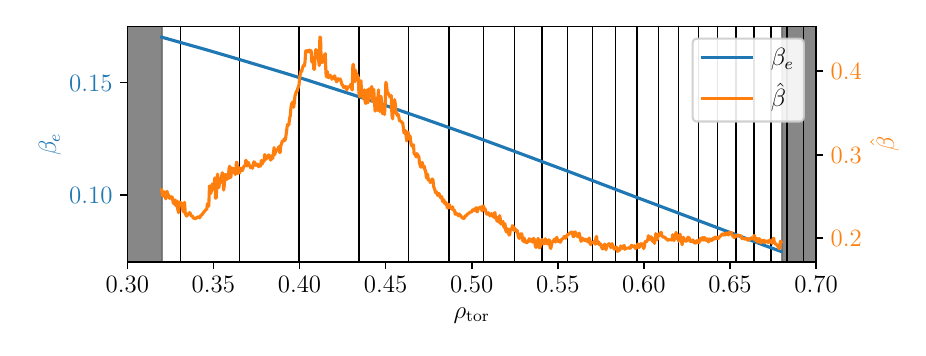}
  \caption{[No evolution of $\dbp$] .
 Real part of the parallel current fluctuation associated with the $n=10$ mode, showing the characteristic current sheet structure concentrated around the rational surface (top). Radial profiles of $\beta_e$ and $\hat{\beta}=\beta_e\big[(a/L_{T_e}+a/L_n)/\hat{s}\big]^2$, illustrating the electromagnetic drive for microtearing and the location of the most favourable surface (bottom). In the bottom panel the rational surfaces are marked by black lines.  
  }
  \label{fig:2d_modestructures_n10_noBpar_jpar}
\end{figure}

Figure~\ref{fig:2d_modestructures_n10_noBpar_jpar} provides additional evidence that the mode obtained in the global simulation without \(\delta \! B_\parallel\) corresponds to an MTM. The upper panel shows the real part of the parallel current fluctuation \(J_\parallel\) associated with the \(n_0=10\) mode as a function of \(\rho_{\mathrm{tor}}\) and \(z\). The fluctuation is strongly localised radially around a single rational surface, where \(k_\parallel \simeq 0\), and exhibits a narrow current sheet structure along the field line. Such localisation of \(J_\parallel\) is a defining characteristic of microtearing modes, which are driven by magnetic reconnection at rational surfaces~\cite{drake1977,Guttenfelder2012,Larakers2021}. The lower panel shows the corresponding radial profiles of the electron beta \(\beta_e\) and the effective microtearing drive parameter \(\hat{\beta} = \beta_e[(a/L_{T_e}+a/L_n)/\hat{s}]^2\). The peak in \(\hat{\beta}\) coincides closely with the radial location of the current sheet observed in the upper panel, indicating that the mode localises where the electromagnetic drive for microtearing is strongest. This alignment between the current localisation and the maximum in the microtearing drive is consistent with theoretical and numerical studies of MTMs, in which the instability selects the most favourable rational surface based on local equilibrium conditions.

Taken together, the presence of a narrow parallel current sheet centred on a rational surface and its alignment with the peak microtearing drive provide strong confirmation that the mode identified in the global calculation is an MTM, rather than a hKBM or ballooning-type instability. The nonlinear consequences of isolating this mode, by removing $\dbp$ from the nonlinear simulations as well as from the linear ones, have been examined in local simulations of this operating point, in which the isolated microtearing turbulence saturates at negligible transport in three independent codes~\cite{giacomin2023b}.

\begin{figure}
  \centering
  \begin{minipage}[t]{0.33\textwidth}
    \centering
    \includegraphics[]{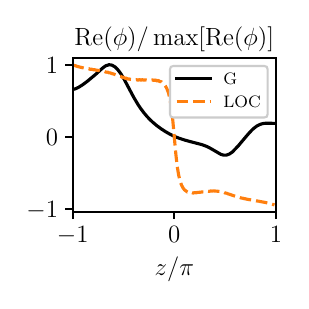}
  \end{minipage}  \begin{minipage}[t]{0.33\textwidth}
    \centering
    \includegraphics[]{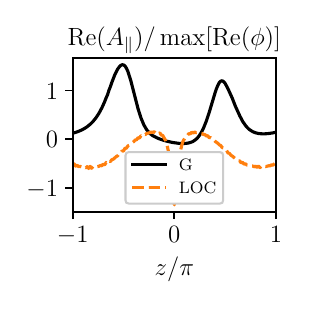} \end{minipage} \hfill
  \caption{[No evolution of $\dbp$] Eigenfunctions of the real part of the fluctuating electrostatic potential (left-hand side), and the parallel magnetic vector potential (right-hand side) for $n_0 = 10$. Eigenfunctions are shown for both the global code (G) and from the local code (LOC). The global eigenfunction is sampled along the ballooning field line passing through the radial position of maximum $|\phi|$, and the local eigenfunction is evaluated at the corresponding flux-surface location so that both represent the same physical radius. 
}
  \label{fig:1d_modestuctures_n10_noBpar}
\end{figure}

The corresponding one-dimensional eigenfunctions are shown in Figure~\ref{fig:1d_modestuctures_n10_noBpar}. The global eigenfunctions are directly compared with the corresponding local flux-tube eigenfunctions evaluated at the same radial location. In both cases, \(\phi\) and \(A_\parallel\) display the expected parity properties of the MTM (opposite parity to the hKBM)\cite{kennedy2023a,patel2021} although the agreement between the local and global codes is much poorer. The comparatively poor agreement between the local and global eigenfunctions obtained for the MTM-like modes is expected and reflects the intrinsic radial localisation of microtearing modes. In global simulations, MTMs are strongly tied to specific rational surfaces, where \(k_\parallel \simeq 0\), and their structure adjusts to the local equilibrium and magnetic shear at that surface. By contrast, in the local flux-tube formulation the equilibrium is frozen at a single reference radius and the same rational surface is effectively assumed to exist everywhere across the perpendicular domain. This removes the distinguished role of a particular flux surface and smears the tearing layer over the numerical box. As a result, even when local and global calculations identify modes with similar frequencies and growth rates, their parallel eigenfunctions need not agree. The discrepancy therefore reflects the breakdown of strict scale separation for MTMs rather than a numerical inconsistency between the two approaches.

\subsection{Linear simulations of short-wavelength instabilities} \label{subsec:short-wavelength}

The strong localisation of the MTM to a single rational surface (Figure~\ref{fig:2d_modestructures_n10_noBpar}) highlights a key challenge for global simulations, which becomes increasingly severe for shorter-wavelength instabilities. Electromagnetic modes are inherently sensitive to alignment with rational surfaces~\cite{Larakers2021}, and their eigenfunctions can be confined to very narrow radial regions between rational surfaces.  The effect is exacerbated further out in radius where $q(\rho_{\mathrm{tor}})$ varies more rapidly and magnetic shear is stronger [see Figure~\ref{fig:profiles}(g) and (h)] which reduces the distance between rational surfaces. Capturing these modes in global GK therefore requires extremely fine radial resolution and careful control of dissipation. This sensitivity also complicates nonlinear studies: insufficient resolution or excessive hyperdiffusion can suppress physically relevant electromagnetic activity, while under-resolved or weakly damped fluctuations can manifest as spurious high-$k_y$ electromagnetic tails in the flux spectra, mimicking residual transport from unresolved MTM-like dynamics (see e.g.,~\cite{Larakers2021}).

\begin{figure}
    \centering
    \includegraphics[]{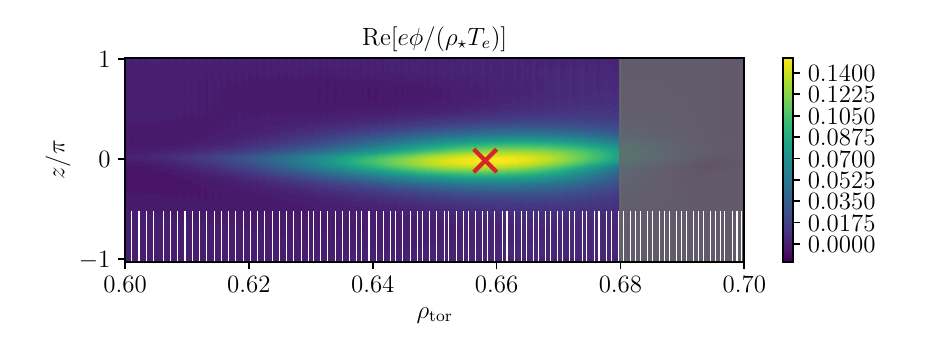}
    \includegraphics[]{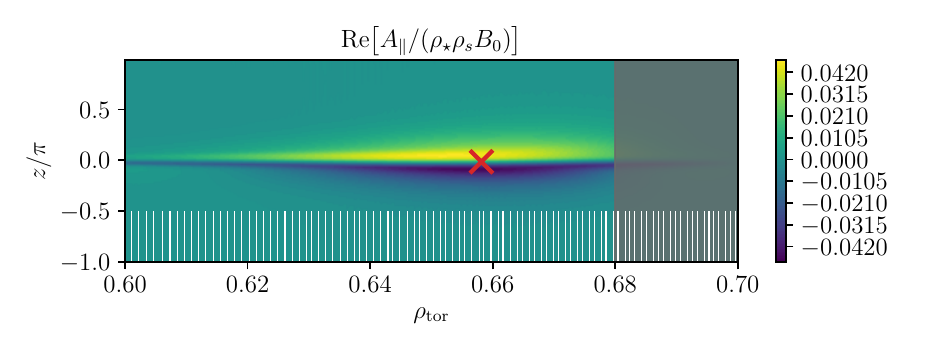}
    \includegraphics[]{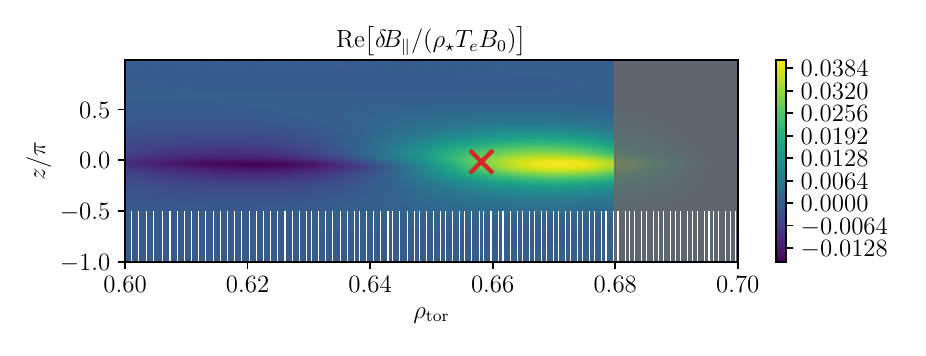}
    \caption{[Long-wavelength solver for $\dbp$] Real part of the fluctuating electrostatic potential (top), parallel magnetic vector potential (middle), and parallel magnetic perturbation (bottom) as functions of the radial coordinate $\rho_{\text{tor}}$ and the field-line-following coordinate $z$ at the final time step of a linear calculation. A global phase factor is applied to all fields such that the complex phase of the maximum electrostatic potential is removed. Results are shown for a mode with toroidal mode-number $n_0=100.$ White lines mark the location of rational surfaces where $q = m/10, \, m \in \mathbb{Z}^{+}$. The buffer regions are shaded grey. The mode peaks close to the edge of the computational box (red cross) where the rational surfaces become very closely spaced [note that the figure is zoomed in on the right edge of the domain].
}
  \label{fig:2d_modestructures_n100}
\end{figure}

Fortunately, broader modes such as the hKBM remain accessible at achievable numerical resolutions. Figure~\ref{fig:2d_modestructures_n100} shows the structure of the \(n_0=100\) mode, plotting the absolute values of \(\phi\), \(A_\parallel\), and \(B_\parallel\) as functions of \(\rho_{\mathrm{tor}}\) and \(z\) from a global simulation employing the long-wavelength \(\delta B_\parallel\) solver. In contrast to the low-\(n_0\) case, the hKBM peaks close to the edge of the computational domain, where the pressure-gradient drive is strongest. In this region, the density of rational surfaces is extremely high and their separation becomes comparable to, or smaller than, the radial grid spacing. As a result, resolving high-\(n_0\) microtearing-like modes, which rely on the precise alignment of the electron diamagnetic drive with individual rational surfaces~\cite{Larakers2021}, would require prohibitively fine radial resolution. This provides a natural explanation for why MTM-like modes are increasingly difficult to capture at high \(n_0\) in global simulations. Recall that although MTMs are subdominant in the middle of the domain, they become dominant deeper into the core of STEP-EC-HD. 

\begin{figure}
  \centering
  \begin{minipage}[t]{0.33\textwidth}
    \centering
    \includegraphics[]{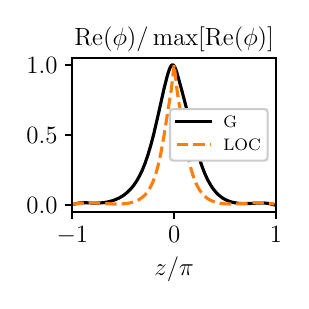}
  \end{minipage}\hfill  \begin{minipage}[t]{0.33\textwidth}
    \centering
    \includegraphics[]{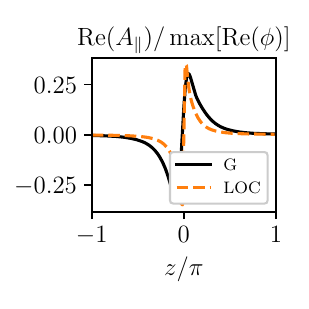} \end{minipage} \hfill
  \begin{minipage}[t]{0.33\textwidth}
    \centering
    \includegraphics[]{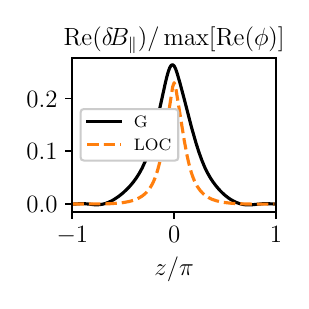}
\end{minipage}
  \caption{[Long-wavelength solver for $\dbp$] Eigenfunctions of the real part of the fluctuating electrostatic potential (left-hand side), parallel magnetic vector potential (middle), and parallel magnetic perturbation (right-hand side) for $n_0 = 100$. Eigenfunctions are shown for both the global code (G) and from the local code (LOC). The global eigenfunction is sampled along the ballooning field line passing through the radial position of maximum $|\phi|$, and the local eigenfunction is evaluated at the corresponding flux-surface location so that both represent the same physical radius. 
}
  \label{fig:1d_modestuctures_n100}
\end{figure}

The hKBM (that drives the most transport in local GK simulations of STEP-EC-HD) does not rely on localisation to a single rational surface, and its broader radial and parallel structure is therefore well resolved at the available resolution. Consistent with this, the global eigenfunctions at \(n_0=100\) exhibit the same qualitative structure and parity as those obtained from corresponding local flux-tube calculations evaluated at the radial location of peak drive, providing further confidence that the long-wavelength global solver is correctly capturing the dominant instability (Figure~\ref{fig:1d_modestuctures_n100}).

\subsection{Constraints on the minimum wavelength}
\label{subsubsec:global_constraints}

As mentioned in Section~\ref{subsec:local_vs_global}, the need to resolve rational surfaces places a limit on the smallest resolvable binormal wavelength. At sufficiently large toroidal mode number \(n_0\), the spacing between adjacent rational surfaces can become comparable to, or smaller than, the radial grid scale. This imposes a fundamental constraint on the maximum mode number that can be meaningfully resolved in global gyrokinetic simulations for a given number of radial grid points\footnote{In fact, this constraint can be thought of as independent of the resolution of the gyrokinetic solver itself as this constraint can also be set by the effective radial resolution of the numerical Grad--Shafranov equilibrium. Numerical equilibrium reconstructions inevitably contain small but finite radial discontinuities in quantities such as the safety factor, magnetic shear, and pressure gradients, which appear below a characteristic radial scale corresponding to the smallest scale on which the equilibrium can be regarded as smooth.}. For high-\(n_0\) electromagnetic modes whose radial structure is controlled by alignment with rational surfaces (e.g., MTMs), the characteristic radial width of the harmonic structure scales as \(\Delta\rho \sim 1/(n_0 q')\), where \(q'=\mathrm{d}q/\mathrm{d}\rho\). 

\begin{figure}
  \centering
  \begin{minipage}[t]{0.5\textwidth}
    \centering
    \includegraphics[width=\textwidth]{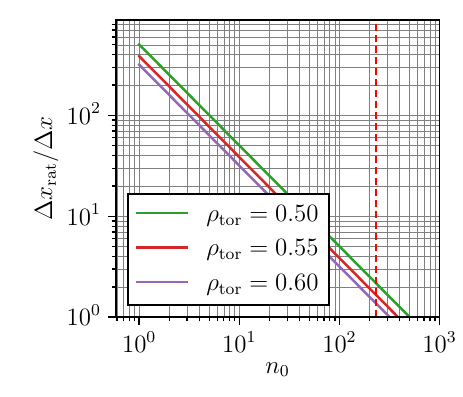}
  \end{minipage}\hfill
  \begin{minipage}[t]{0.5\textwidth}
    \centering
    \includegraphics[]{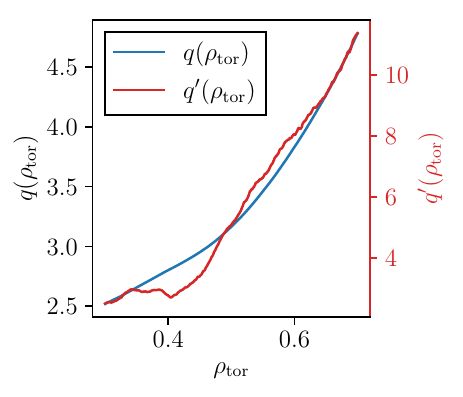}
\end{minipage}
  \caption{Number of radial grid cells per rational-surface spacing,
$\Delta x_{\rm rat}/\Delta x = 1/(n_0 |q'(x)| \Delta x)$, as a function of toroidal mode number $n_0$
at several fixed radial locations $\rho_{\mathrm{tor}}=0.4$--$0.6$ (left-hand side).
Safety-factor profile $q(\rho_{\mathrm{tor}})$ and magnetic shear $q'(\rho_{\rm tor})=\mathrm{d}q/\mathrm{d}\rho_{\rm{tor}}$ across the radial domain (right-hand-side). 
The loss of resolution at large $n_0$ is controlled by the spacing of rational surfaces rather than
by the radial width of the mode envelope.
}
  \label{fig:global_rational_spacing}
\end{figure}

Figure~\ref{fig:global_rational_spacing} shows the number of radial grid cells per rational-surface spacing as a function of $n_0$ (left-hand side). This diagnostic provides a quantitative criterion for numerical adequacy.
For $n_0 \lesssim 200$, the rational-surface spacing is resolved by several grid
points throughout the domain.
At larger $n_0$, the number of grid points per rational spacing drops below
$\sim 3-4$, at which point individual rational surfaces are no longer resolved and
modes become susceptible to grid locking or artificial radial pinning.
Importantly, this constraint is far more restrictive than the resolution requirement associated
with the overall mode envelope (e.g.\ its full-width at half-maximum), which remains well resolved
over the same range of $n_0$. The problem is further exacerbated towards the edge of the domain where the magnetic shear is larger [see Figure~\ref{fig:global_rational_spacing} (right-hand side)]. 

This constraint is particularly restrictive for microtearing modes, which rely on precise alignment of the electron diamagnetic drive with individual rational surfaces and therefore possess intrinsically narrow radial structure. By contrast, broader instabilities such as the hKBM have radial widths that comfortably exceed \(\rho_\star\) and are typically less sensitive to the grid resolution. As a result, global simulations can robustly capture large-scale, pressure-gradient-driven modes at accessible resolutions, while simultaneously struggling to resolve fine-scale microtearing physics at high toroidal mode number.

More generally, this reflects a fundamental multiscale challenge that is common to both global and local gyrokinetic formulations. Capturing global hKBM dynamics requires a sufficiently large radial domain, whereas resolving MTMs demands extremely fine radial resolution and a highly accurate equilibrium representation. Achieving both simultaneously remains prohibitively expensive with current numerical approaches. In this work, we therefore focus on resolving the structure of the dominant hKBM modes in global simulations, for which the relevant spatial scales are well separated from both the high order rational surface structure and equilibrium fine structure. The inability to simultaneously resolve very high-\(k_y\) microtearing tails does not affect the identification or characterisation of the primary pressure-gradient-driven instabilities of interest. An analogous limitation is evident in local simulations, where studies have been unable to simultaneously resolve hKBM and MTM turbulence owing to similar multiscale constraints, and MTM-only simulations (made possible by artificially suppressing the $\dbp$ hKBM drive) typically exhibit very low transport levels~\cite{giacomin2023b}. Again we emphasise that MTMs are the dominant mode in the core (shallow gradient region)~\cite{giacomin2024b}. The scale of the difficulty is set by the parallel current sheet of Figure~\ref{fig:2d_modestructures_n10_noBpar_jpar}, which is $\Delta\rho_{\mathrm{tor}} \approx 1.6\times10^{-3}$ wide at half maximum, about $0.55\,\rho_i$, whereas the radial spacing of a global domain wide enough to contain the hKBM turbulence is $\Delta\rho_{\mathrm{tor}} \sim 3.1\times10^{-3}$, so the layers on which the microtearing mode is centred fall below the grid of any global simulation that can also hold the transport-relevant dynamics. A systematic treatment of core MTM transport in local and global simulations therefore remains a future extension to this work.

\section{Global nonlinear simulations of electromagnetic turbulence in STEP-EC-HD} \label{sec:nonlinear}

It was shown in Section~\ref{sec:linear} that the long-wavelength $\dbp$ solver in global \texttt{GENE} captures radially extended hKBM instabilities in STEP-EC-HD. The nonlinear calculations that follow explore hKBM-driven turbulence in the global GK framework and the importance of global effects on the nonlinear dynamics.

Local nonlinear simulations of STEP-EC-HD return fluxes far in excess of the available sources at the nominal $\beta_e$~\cite{giacomin2023b,kennedy2024}. Reference~\cite{Kennedy2025EM} (and references therein) extends this to a systematic scan of over one hundred flux-tube simulations of this operating point, performed on the mid-radius $q = 3.5$ surface ($\Psi_n = 0.49$, i.e.\ the $\rho_{\mathrm{tor}} = 0.55$ surface of Table~\ref{tab:local_params}), in which $\beta^\prime$ is scaled self-consistently with $\beta_e$ (i.e., equivalent to scaling the pressure profile). Two distinct transitions were identified. Increasing $q^2 \beta_e$ through a critical value triggers the transition to very large fluxes. Importantly, it was shown that the location of this threshold can be predicted by applying the zonal-flow torque-competition theory of~\cite{Zhang2025a,Zhang2025b} to STEP-EC-HD. Increasing $q^2\beta_e$ further produces a reverse transition, in which $\beta^\prime$ stabilisation of the hKBM restores a moderate-flux regime, corresponding to access to second stability. The nominal operating point lies above the first of these thresholds (but below the second-stability window) where local GK returns very large fluxes. Whether finite-$\rho_\star$ physics changes that conclusion is addressed here using global simulations.

\subsection{Nonlinear global GK turbulence in STEP-EC-HD}
\label{subsec:nonlinear_GK_STEP-EC-HD}

\begin{table}[h!]
\caption{Local gyrokinetic parameters on the five flux surfaces of the local (LOC) simulations of this section; the entries are the values \texttt{GENE} used. Here $\beta^\prime = -\beta\,(a/L_p)$ with $\beta$ the total thermal beta and $L_p$ the pressure gradient scale length, $\rho_s = \sqrt{T_e m_i}/(eB_{\mathrm{ref}})$, $c_s = \sqrt{T_e/m_i}$, $\nu_{ee}$ is the electron-electron collision frequency, and $a = 2.01$~m is the half-diameter of the last closed flux surface at the equatorial midplane. These normalisations are not the ones used in~\cite{giacomin2023b}, where gradients are taken with respect to $r/a$ rather than $\rho_{\mathrm{tor}}$ and $\beta_e$ and $\rho_s$ are referred to the field at the centre of the flux surface rather than to $B_{\mathrm{ref}}$, so the entries here differ from those tabulated there for the same equilibrium.}
\label{tab:local_params}
\centering
\renewcommand{\arraystretch}{1.3}
\begin{tabular}{@{}lccccc@{}}
\toprule
$\rho_{\mathrm{tor}}$ & 0.40 & 0.45 & 0.50 & 0.55 & 0.60 \\
\midrule
$\Psi_n$ & 0.28 & 0.35 & 0.42 & 0.49 & 0.56 \\
$q$ & 2.80 & 2.95 & 3.16 & 3.46 & 3.83 \\
$\hat{s}$ & 0.39 & 0.52 & 0.80 & 1.06 & 1.26 \\
$\beta_e$ & 0.152 & 0.140 & 0.126 & 0.112 & 0.098 \\
$\beta^\prime$ & $-0.56$ & $-0.59$ & $-0.61$ & $-0.61$ & $-0.60$ \\
$\rho_s$ [mm] & 6.28 & 6.07 & 5.86 & 5.64 & 5.41 \\
$T_e$ [keV] & 12.9 & 12.0 & 11.2 & 10.4 & 9.6 \\
$n_e$ [$10^{20}\,\mathrm{m^{-3}}$] & 1.99 & 1.96 & 1.90 & 1.83 & 1.72 \\
$T_i/T_e$ & 1.09 & 1.07 & 1.05 & 1.03 & 1.02 \\
$n_D/n_e$ & 1.00 & 1.00 & 1.00 & 1.00 & 1.00 \\
$a/L_{n_e}$ & 0.29 & 0.45 & 0.69 & 0.97 & 1.30 \\
$a/L_{T_e}$ & 1.28 & 1.39 & 1.49 & 1.57 & 1.64 \\
$a/L_{T_i}$ & 1.67 & 1.77 & 1.82 & 1.84 & 1.85 \\
$(a/c_s)\nu_{ee}$ & 0.023 & 0.026 & 0.029 & 0.032 & 0.035 \\
\bottomrule
\end{tabular}
\end{table}

\begin{figure}
  \centering
  \begin{minipage}[t]{0.5\textwidth}
    \centering
    \includegraphics[]{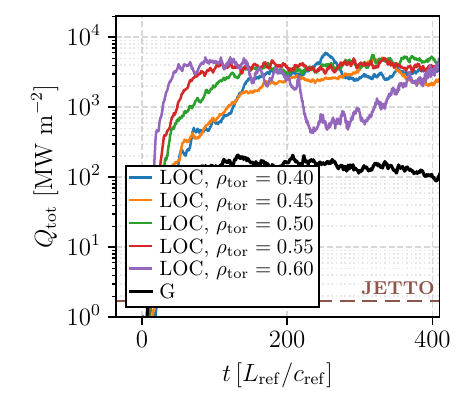}
  \end{minipage}\hfill
  \begin{minipage}[t]{0.5\textwidth}
    \centering
    \includegraphics[]{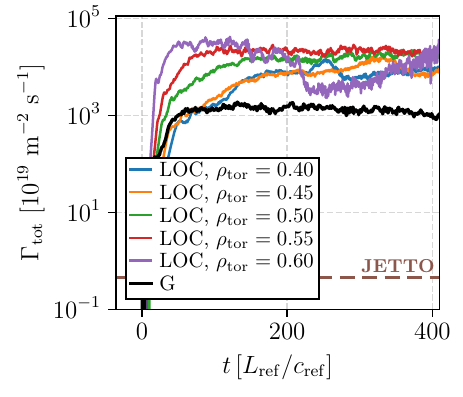}
\end{minipage}
  \caption{Time traces of the volume-averaged total heat (left-hand side) and particle (right-hand side) fluxes from nonlinear simulations of STEP-EC-HD. Results are shown for both a global calculation (G) and for local calculations (LOC) on different flux surfaces. The global calculation uses the long-wavelength $\dbp$ solver of \cite{sheffield2024}. The local calculations evolve all fields fully self-consistently. Brown dashed lines mark balance with the available sources, computed from the SPR-045 JETTO run underlying~\cite{tholerus2024} as in Table~\ref{tab:flux_comparison} (the sources integrated within each flux surface and divided by its area): the panels use the mid-radius values, $1.7\,\mathrm{MW\,m^{-2}}$ and $0.45\times10^{19}\,\mathrm{m^{-2}\,s^{-1}}$.
}
  \label{fig:local_global_nonlinear_comparison}
\end{figure}

Figure~\ref{fig:local_global_nonlinear_comparison} shows time traces of the total heat and particle fluxes returned by local and global nonlinear GK simulations of STEP-EC-HD on the grids of Table~\ref{tab:resolution}. The local (LOC) calculations of this section are carried out on five flux surfaces, $\rho_{\mathrm{tor}} = 0.40$ to $0.60$, spanning the interior of the global domain, with their local gyrokinetic parameters collected in Table~\ref{tab:local_params}. After the initial linear growth phase the fluxes of both frameworks are typically more than two orders of magnitude larger than the levels compatible with the available STEP sources~\cite{giacomin2023b}, and global effects do not prevent the transition to non-zonal, electromagnetically dominated states carrying extreme transport\footnote{We remind the reader that these simulations are gradient driven, the initial profiles held near their targets by the numerical sources of Section~\ref{subsec:boundary_conditions_numerical_sources}. In a flux-driven simulation the profiles would instead relax towards gyrokinetic power balance~\cite{giacomin2024b}, which would eliminate this extreme-transport state. \ref{app:sources} examines a related relaxation by switching the sources off, and finds the transport still an order of magnitude above the available heating.}. The local traces show intermittent excursions lasting several tens of $L_{\mathrm{ref}}/c_{\mathrm{ref}}$, the clearest at $\rho_{\mathrm{tor}} = 0.60$, where the local total heat flux falls from approximately $3800$ to $550\,\mathrm{MW\,m^{-2}}$ for roughly a hundred $L_{\mathrm{ref}}/c_{\mathrm{ref}}$ before recovering. These excursions coincide with the long-wavelength non-zonal amplitudes transiently overtaking the zonal component~\cite{giacomin2023b}. Saturated flux levels are quoted throughout as averages over specified time windows.

\begin{figure}
  \centering
  \begin{minipage}[t]{0.5\textwidth}
    \centering
    \includegraphics[]{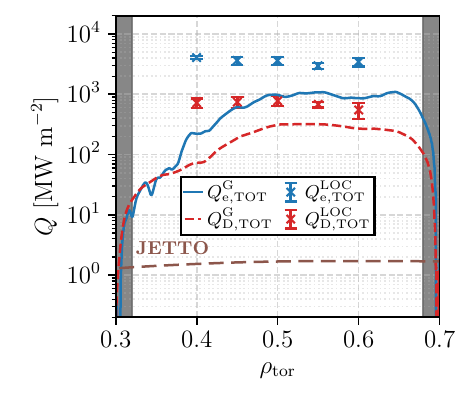}
  \end{minipage}\hfill
  \begin{minipage}[t]{0.5\textwidth}
    \centering
    \includegraphics{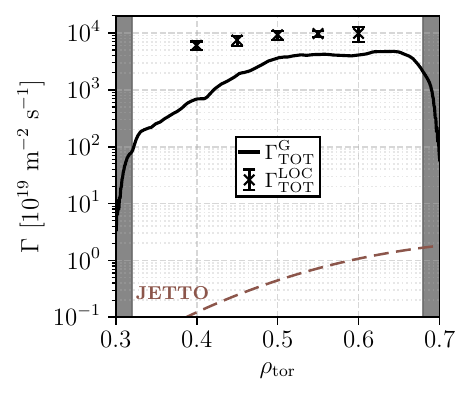}
\end{minipage}
  \caption{Time-averaged (over the last 10\% of the simulation) heat (left-hand side) and particle  (right-hand side) fluxes as functions of the radial coordinate. Brown dashed lines mark balance with the available sources, computed from the SPR-045 JETTO run underlying~\cite{tholerus2024} as in Table~\ref{tab:flux_comparison} (the sources integrated within each flux surface and divided by its area): the panels use the radially resolved profiles, which vary with radius.
}
  \label{fig:local_global_nonlinear_comparison_timeaveraged}
\end{figure}

Figure~\ref{fig:local_global_nonlinear_comparison_timeaveraged} shows the time-averaged heat and particle fluxes at late times as functions of the radial coordinate. The global fluxes increase with radius across most of the domain, following the radial variation of the instability drive, whereas the local heat flux is almost independent of radius. Note that the local fluxes exceed the global fluxes at every radius. This is likely because the buffer regions and Dirichlet boundary conditions of the global domain artificially constrain the largest radial scales, which the flux tube allows to develop freely; indeed, local simulations of this operating point are known to develop radially extended turbulent structures in the absence of equilibrium flow shear~\cite{giacomin2023b}. Nevertheless, the two frameworks agree to within an order of magnitude over $\rho_{\mathrm{tor}} \ge 0.45$, with heat fluxes differing by factors of about $3$ to $5$. The disparity is largest at $\rho_{\mathrm{tor}} = 0.40$, where the global flux is smallest and the influence of the buffer region is strongest. Table~\ref{tab:flux_comparison} compares both sets of fluxes with the levels in balance with the available heating and particle sources at the radii of the local analysis.

\begin{table}[h!]
\caption{Total heat flux $Q_{\mathrm{tot}}$ (electrons plus ions, electrostatic plus electromagnetic) and electron particle flux $\Gamma_{\mathrm{e}}$ from the local (LOC) and global (G) simulations at the radii of the local analysis, averaged over the final $10$\% of each run, together with the levels in balance with the available sources (JETTO). The latter are the heating and fuelling of the SPR-045 JETTO run underlying~\cite{tholerus2024}, integrated within each surface and divided by its area, with no radiative losses subtracted.}
\label{tab:flux_comparison}
\centering
\renewcommand{\arraystretch}{1.3}
\begin{tabular}{@{}lccccc@{}}
\toprule
$\rho_{\mathrm{tor}}$ & 0.40 & 0.45 & 0.50 & 0.55 & 0.60 \\
\midrule
$Q_{\mathrm{tot}}$ (LOC) [$\mathrm{MW\,m^{-2}}$] & 4780 & 4380 & 4390 & 3650 & 4010 \\
$Q_{\mathrm{tot}}$ (G) [$\mathrm{MW\,m^{-2}}$] & 295 & 800 & 1290 & 1400 & 1130 \\
$Q_{\mathrm{tot}}$ (JETTO) [$\mathrm{MW\,m^{-2}}$] & 1.5 & 1.6 & 1.7 & 1.7 & 1.7 \\
\midrule
$\Gamma_{\mathrm{e}}$ (LOC) [$10^{19}\,\mathrm{m^{-2}\,s^{-1}}$] & 6050 & 7500 & 9300 & 9750 & 9850 \\
$\Gamma_{\mathrm{e}}$ (G) [$10^{19}\,\mathrm{m^{-2}\,s^{-1}}$] & 690 & 1830 & 3640 & 4180 & 4100 \\
$\Gamma_{\mathrm{e}}$ (JETTO) [$10^{19}\,\mathrm{m^{-2}\,s^{-1}}$] & 0.12 & 0.25 & 0.45 & 0.73 & 1.08 \\
\bottomrule
\end{tabular}
\end{table}

\begin{figure}
  \centering
  \begin{minipage}[t]{0.5\textwidth}
    \centering
    \includegraphics[]{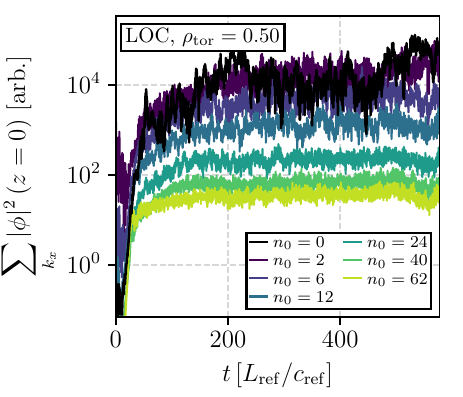}
  \end{minipage}\hfill
  \begin{minipage}[t]{0.5\textwidth}
    \centering
    \includegraphics[]{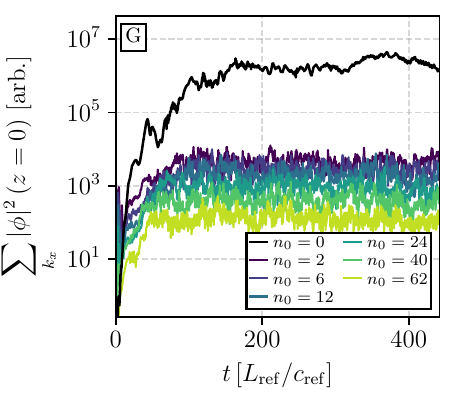}
\end{minipage}\\[6pt]
  \includegraphics[]{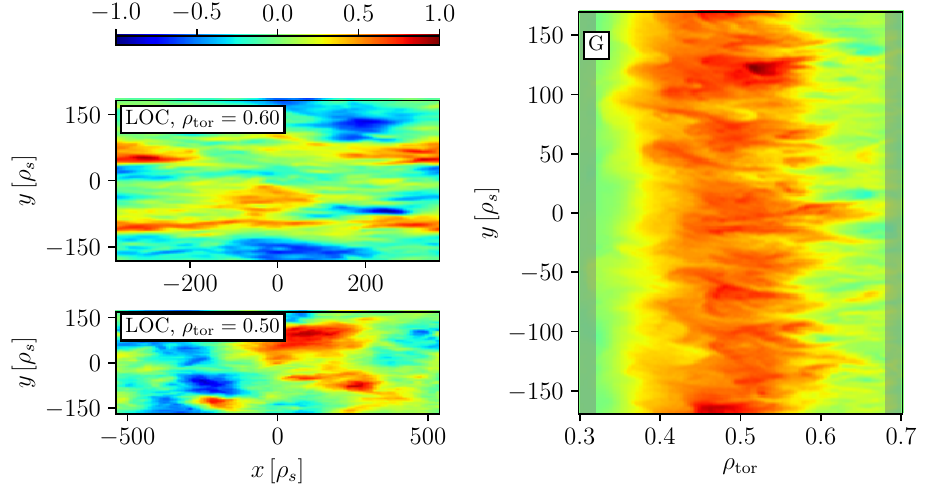}
  \caption{Top: time traces of the squared electrostatic potential amplitude at the outboard midplane, $\sum_{k_x}|\phi|^{2}(z=0)$, for toroidal mode numbers $n_0 = 2$ to $62$ (colours) and the zonal component $n_0 = 0$ (black), in each code's internal units, for the local (LOC) simulation at $\rho_{\mathrm{tor}} = 0.50$ (left-hand side) and the nominal global simulation (G) (right-hand side). Bottom: the electrostatic potential in the perpendicular plane at the outboard midplane at the end of each simulation, for the flux tubes at $\rho_{\mathrm{tor}} = 0.60$ and $0.50$ and for the global simulation, each panel normalised to its maximum amplitude. All panels are drawn at the true aspect ratio of their boxes, whose radial extents differ: the flux-tube box dimensions follow from the safety factor and magnetic shear at each radius through the standard flux-tube quantisation, while the global domain is the physical annulus. Grey shading marks the buffer regions of the global simulation.}
  \label{fig:phi2_by_ky}
\end{figure}

Figure~\ref{fig:phi2_by_ky} follows the saturation mode by mode, showing the squared electrostatic potential amplitude at the outboard midplane against time for a selection of toroidal mode numbers, in the local simulation at $\rho_{\mathrm{tor}} = 0.50$ and in the global simulation. The two frameworks treat the largest scales differently. In the local simulation the lowest mode numbers grow throughout, with $n_0 = 2$ still growing at the end of the simulation, and the non-zonal amplitudes repeatedly overtake the zonal component, in about half of the late-time snapshots. In the global simulation every non-zonal mode is statistically stationary after $t \approx 100\,L_{\mathrm{ref}}/c_{\mathrm{ref}}$ and the zonal potential stands nearly two orders of magnitude above the summed non-zonal amplitudes, consistent with the Dirichlet boundary condition constraining the largest radial scales, as discussed above. The global simulation accesses high flux states with a strong stochastic magnetic field contribution to the transport (\ref{app:lowbeta}), but notably retains a dominant zonal potential throughout the simulation. In the flux tube, the zonal potential is considerably weaker, and the transport fluxes are larger. The lower row of Figure~\ref{fig:phi2_by_ky} shows the potential in the perpendicular plane at the outboard midplane at the end of each simulation, for the flux tubes at $\rho_{\mathrm{tor}} = 0.60$ and $0.50$ and for the global simulation. At both radii the flux-tube potential is organised into a few radially elongated structures, several hundred $\rho_s$ in radial extent, whereas the global potential is dominated by its zonal component and the eddies superposed on it are broken up radially on much shorter scales. The radial extents in local and global simulations are not comparable parameters. The global domain is the physical annulus, its width in $\rho_s$ determined by $\rho_\star$, while the flux-tube box width is a free parameter, set generously so that the saturation level is not imposed by the box. The global description of the large-flux state is affected by a box-scale zonal potential that is effectively set by the Dirichlet boundary conditions, while this is not imposed in the flux-tube.

\begin{figure}
  \centering
\includegraphics[]{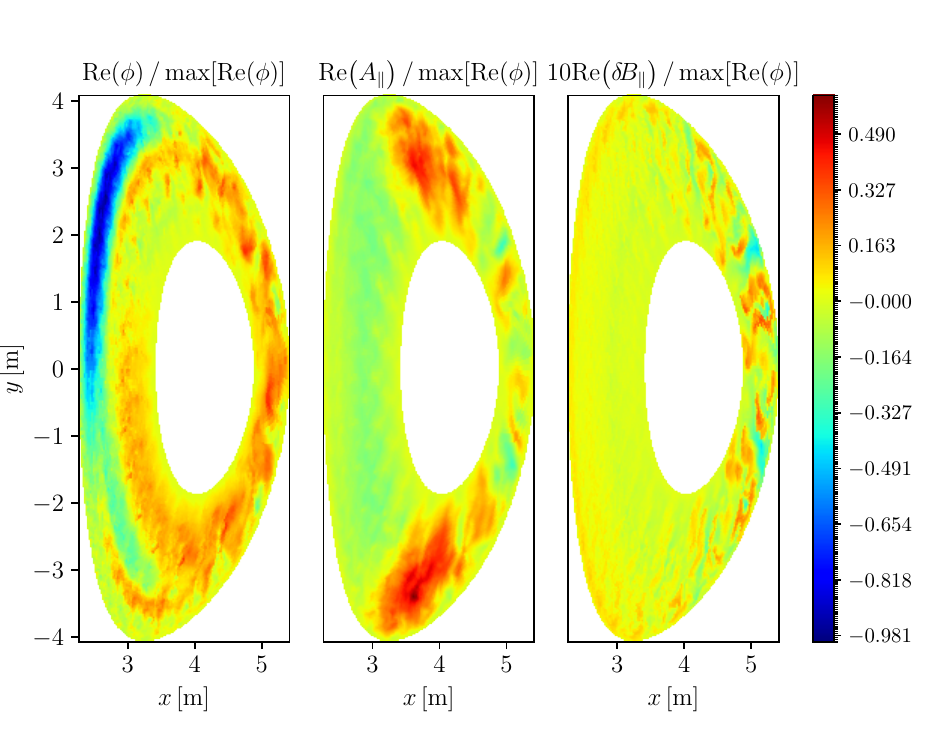}
  \caption{Poloidal cross sections of the real part of the fluctuating electrostatic potential (left-hand side), parallel magnetic vector potential (middle), and parallel magnetic perturbation (right-hand side)  obtained from a global nonlinear electromagnetic gyrokinetic simulation of STEP-EC-HD.  Note that
the aspect ratio of the panels corresponds to that of the STEP-EC-HD cross section.
}
  \label{fig:phi_apar_torus}
\end{figure}

Figure~\ref{fig:phi_apar_torus} presents contour plots in the poloidal plane of the electrostatic potential $\phi$, the parallel vector potential $A_{\parallel}$ and $\dbp$ from the global nonlinear simulation. The electrostatic potential (left-hand side) is weakly ballooning towards the outboard midplane, where curvature and pressure-gradient drive are strongest, and displays even parity about the midplane, characteristic of drift-wave or pressure-gradient-driven turbulence. The parallel vector potential (middle) exhibits an odd-parity structure with lobes above and below the midplane, indicating Alfv\'{e}nic magnetic perturbations with a twisting parity. The large amplitudes of both fields indicate strong electromagnetic coupling, consistent with an Alfv\'{e}nic instability like the hKBM. The perturbations in $\dbp$ (right-hand side) are small, even though its inclusion is essential for capturing the hKBM in linear simulations. The nonlinear dynamics in STEP-EC-HD are therefore of mixed drift-Alfv\'{e}nic character, with pressure-gradient-driven turbulence and magnetic-field-line bending in competition.

\begin{figure}
  \centering
  \begin{minipage}[t]{0.5\textwidth}
    \centering
    \includegraphics[]{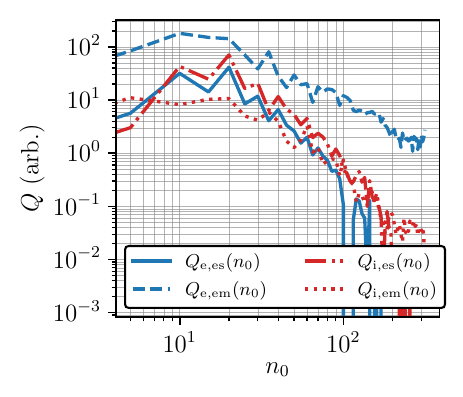}
  \end{minipage}\hfill
  \begin{minipage}[t]{0.5\textwidth}
    \centering
    \includegraphics[]{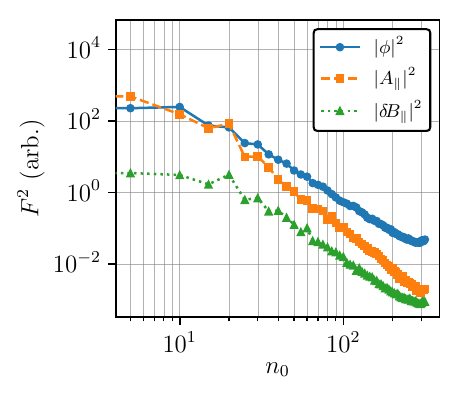}
\end{minipage}
  \caption{Spectral characteristics of turbulent transport and field amplitudes. Left-hand side: Ion and electron heat-flux spectra \(Q_s(n_{0})\) decomposed into electrostatic (es) and electromagnetic (em) contributions as functions of the toroidal mode number \(n_0\). Right-hand side: Fluctuation energy spectra \(F^2\) of the electrostatic potential \(|\phi|^2\), parallel vector potential \(|A_\parallel|^2\), and parallel magnetic field \(|\delta \!B_\parallel|^2\).  }
  \label{fig:flux_spectra}
\end{figure}

Figure~\ref{fig:flux_spectra} shows the electrostatic and electromagnetic contributions to the volume-averaged ion and electron heat flux spectra as functions of the toroidal mode number $n_0$. Relatively low-$k_y$ modes make the largest contribution to the total heat flux, though a pronounced tail persists in the electromagnetic electron heat flux at higher $k_y$, similar to that observed in earlier local simulations~\cite{giacomin2023b} but more substantial in the present STEP-EC-HD case. MTMs are not well resolved in these simulations (see Section~\ref{subsec:short-wavelength}), so are unlikely to explain the high-$k_y$ electromagnetic flux.

Local studies of this operating point have addressed the question of whether MTMs contribute materially to the nonlinear transport. Excluding $\dbp$ linearly stabilises the hKBM, leaving the MTM as the only remaining instability. Dedicated multi-code simulations of the resulting isolated MTM turbulence found essentially negligible transport fluxes at this surface~\cite{giacomin2023b}. Conversely, when the effects of $\dbp$ are approximately retained (through modifying the magnetic drifts without including the field explicitly), the hKBM and a transition to very large fluxes returns~\cite{kennedy2024}. Nonlinear simulations (retaining all three fields) at the innermost analysed surface of this operating point, where the MTM is linearly dominant, similarly find that the MTM-range contribution saturates at a level of order the gyro-Bohm flux with negligible particle transport, with the large fluxes there carried by the weakly unstable long-wavelength hKBM~\cite{giacomin2024b}. These local results are consistent with the large fluxes reported here being driven by the hKBM rather than by subdominant MTM activity. We do not attempt a global repetition of these tests here, as the radial grid required to resolve the current sheets on which the MTM is centred is computationally prohibitive on a domain wide enough to contain the hKBM turbulence (see Section~\ref{subsubsec:global_constraints} and Figure~\ref{fig:global_rational_spacing}). We note, however, that these current sheets are only about half an ion gyroradius wide ($\Delta\rho_{\mathrm{tor}} \approx 1.6\times10^{-3}$), so the equilibrium profiles are essentially constant across them and the local description of the MTM itself is expected to remain adequate.

\subsection{Implications for STEP-EC-HD burning flat top}
\label{subsec:feasibility}

The main purpose of this work is to assess whether global effects impact significantly
on the large fluxes reported from gradient-driven local simulations for STEP-EC-HD in \cite{giacomin2023b}.  Importantly Table~\ref{tab:flux_comparison} shows that local and global gyrokinetic calculations predict fluxes at the reference design gradients that exceed the available sources by more than two orders of magnitude over most of the radial domain (with larger flux predictions from the local runs). The new global results therefore confirm the previous inference from local GK simulations in \cite{giacomin2024b}, that the STEP-EC-HD reference burning flat-top operating point, where confinement was modelled using simplified empirical transport models~\cite{tholerus2024}, is not in a transport steady state.

Gradient-driven calculations are extremely valuable, but cannot address feasibility which requires a flux-driven approach; this is presently impractical for STEP-EC-HD using global nonlinear gyrokinetic simulations.  A physics based flux-driven approach was developed in \cite{giacomin2024b}, where a novel quasilinear reduced transport model was constructed to describe hybrid-KBM turbulence in STEP-like plasmas.  This model was applied to transport calculations for STEP-EC-HD using the T3D code.  In spite of the enormous local anomalous fluxes at the reference operating point (see Fig 16 of \cite{giacomin2024b}), T3D calculations nevertheless suggest that high performance plasma states with fusion power comparable to the design value and more acceptable transport may exist for STEP-EC-HD; this is enabled by $\beta^\prime$ stabilisation~\cite{giacomin2024b}. This was an important finding, though  access to the favourable state was not demonstrated. (We note that a number of further important improvements are required to improve the fidelity and robustness of such calculations.\footnote{These include addressing the impacts of fast alphas and impurities on turbulent transport, evolving the equilibrium self-consistently, and modelling rotation.}) The global gyrokinetic calculations in this paper provide invaluable support for this approach.

The robustness of our above findings using the global framework, have been tested through a number of additional calculations. We have explored the impact of reducing $\beta_e$ at fixed geometry; this is reported in \ref{app:lowbeta}, where the reduced-$\beta_e$ case recovers a moderate-flux, zonally regulated state. We have explored the impact of equilibrium flow shear, at levels from the diamagnetic rotation the device generates for itself up to the quench-rule level; this is reported in \ref{app:flowshear}. We have also examined the relaxation of the profiles when the interior numerical sources are removed; this is reported in \ref{app:sources}.

\section{Conclusions}
\label{sec:conclusions}

Global gyrokinetic simulations retaining the full electromagnetic response, including $\delta \! B_{\parallel}$, have been performed for the conceptual burning flat-top operating point of STEP, STEP-EC-HD~\cite{tholerus2024}, to study the influence of non-local effects on the nonlinear saturation and transport characteristics of electromagnetic turbulence at finite~$\beta$.\footnote{This first global gyrokinetic analysis for STEP includes the impacts of thermal electrons and deuterium, but neglects the possible impacts of impurities and fast particles.} A long-wavelength $\dbp$ solver~\cite{sheffield2024}, recently implemented in \texttt{GENE}, was used to include fluctuations parallel to the magnetic field which were found in previous work~\cite{kennedy2023a,giacomin2023b,kennedy2024} to be essential for capturing the linearly dominant hKBMs in STEP-EC-HD. 

It was shown that the long-wavelength field solver is sufficient to accurately capture the linear mode spectrum in global simulations of STEP-EC-HD. A detailed comparison between a single global calculation, which naturally samples the radial variation of the linear drive, and an ensemble of local flux-tube simulations found close agreement in the mode frequencies, growth rates and eigenfunction structure of the hKBMs previously identified in~\cite{kennedy2023a}. Capturing microtearing modes (MTMs), however, remains more challenging due to their shorter radial wavelengths and the associated stringent numerical resolution requirements.

Importantly, nonlinear global simulations reveal clear evidence of a transition to a state with extremely large heat fluxes close to mid-radius, consistent with the loss of zonal-flow regulation predicted for this operating point at high $q^2\beta_e$ \cite{Kennedy2025EM} and in agreement with previous results from local nonlinear simulations~\cite{giacomin2023b,kennedy2024}. A deliberately simple fixed-geometry global calculation in which $\beta_e$ is reduced by a factor of six (\ref{app:lowbeta}) instead saturates in a moderate-flux, zonally regulated state, placing the nominal operating point above the transition, consistent with the local threshold of~\cite{Kennedy2025EM}; Poincar\'e sections indicate that magnetic-field-line stochasticisation plays an important role in the high-flux state, weakening the impact of zonal flows on turbulence suppression. The transition to high flux states in our global simulations closely resembles findings from local electromagnetic studies (that include $A_{\parallel}$ but not $B_{\parallel}$) of conventional aspect ratio CBC plasmas \cite{Pueschel2013,Zhang2025b}.  We note, however, that the large local fluxes for CBC were not reproduced in the global gyrokinetic simulations of \cite{ishizawa2019,masui2022}, which may be due to features of the adopted model: (i) linear gradient drive terms in these gradient driven simulations are radially localised in the centre of the domain, and vanishing Dirichlet boundary conditions are imposed at the inner and outer radial boundaries, both of which act to limit the radial extent of fluctuations; and (ii) reduced ion-to-electron mass ratios are used in these simulations ($m_i/m_e = 100$~\cite{ishizawa2019} and $400$ in~\cite{masui2022}), which enhances the effective electron collisionless skin-depth and reduces the strength of perpendicular magnetic fluctuations.
Finally we note that global corrections scale with $\rho_{\ast}$, so should be weaker at mid-radius in STEP-EC-HD where $\rho_{\ast} \sim 1.5\times 10^{-3}$, than in CBC where $\rho_{\ast} \sim 0.01$. We have also begun to explore the impact of these differences directly in global simulations; those tests are still maturing and will be reported separately.

The transition to large fluxes in global simulations is robust to the tests applied here. Removing the interior profile control leaves the transport an order of magnitude above the available heating, with the profiles degrading substantially to reach even that level, so the transition to large fluxes is not entirely an artefact of that control (\ref{app:sources}). Global simulations with equilibrium flow shear at the diamagnetic level show little impact on the transport, but at the level of the quench-rule (which would likely require a momentum source) the domain-averaged electromagnetic electron heat flux is substantially reduced but still above the available heating power (\ref{app:flowshear}).

We undertook these global simulations to test whether the very large fluxes previously reported from local gyrokinetic simulations for STEP reference plasma equilibria \cite{giacomin2023b}, are a genuine property of the plasma, or are an artefact of the local limit.  Finite-$\rho_\star$ physics included in these global simulations is found to leave both the existence of the high-flux state at the STEP-EC-HD reference gradients, and the location of the transition threshold in $q^2 \beta_e$, seemingly unchanged. At the STEP-EC-HD reference gradients, in the centre of the radial  domain the predicted transport greatly exceeds the heating power assumed to be available in the integrated modelling; i.e. the transport assumed for this specific operating point is not reproduced by our high-fidelity global gyrokinetic simulation. Whether microtearing turbulence, found to carry negligible transport at mid-radius in STEP-EC-HD in local simulations with multiple codes~\cite{giacomin2023b}, remains as benign in the global framework is left open, since it is computationally prohibitive to simultaneously resolve microtearing current sheets and hKBM turbulence on the global domain. 

The global results presented here are significant, but they are not conclusive from the point of view of STEP.  What matters going forward is: (i) whether transport-consistent plasma equilibria exist that can achieve the target fusion performance (as \cite{giacomin2024b} suggests may be possible through $\beta^{\prime}$ stabilisation~\cite{bourdelle2003}); and (ii) whether such states can be accessed.  These urgent open questions are being actively pursued, but they are also extremely challenging and will require extensions over previous works to include:  self-consistent equilibrium evolution (critical in bootstrap-current dominated plasmas); interactions between turbulence and fast-$\alpha$ particles and impurities; momentum transport; and intrinsic rotation.
The global electromagnetic framework demonstrated in this work will contribute to these extensions, and provide a tool that can assist in the search for a gyrokinetically consistent STEP operating point.

\section*{Acknowledgements}

This work has been
funded by STEP, a major technology and infrastructure programme led by UK Industrial Fusion Solutions Ltd (UKIFS),
which aims to deliver the UK's prototype fusion powerplant
and a path to the commercial viability of fusion. Part of this work was performed using resources provided by the PITAGORA supercomputer from the National Supercomputing
Consortium CINECA, under the projects GEMST24 and GEMST25.  Part of this work was performed using resources provided by the Cambridge Service for Data Driven Discovery (CSD3) operated by the University of Cambridge Research Computing Service (\url{www.csd3.cam.ac.uk}), provided by Dell EMC and Intel using Tier-2 funding from the Engineering and Physical Sciences Research Council (capital grant EP/T022159/1), and DiRAC funding from the Science and Technology Facilities Council (\url{www.dirac.ac.uk}).

\appendix

\section{Comparing local and global eigenfunctions} \label{app:eigenfunction_comparison}

Linear gyrokinetic eigenfunctions are defined only up to an arbitrary complex phase. 
If $\boldsymbol{\Psi}(z) = (\phi, A_{\parallel}, B_{\parallel})$ is an eigenmode, then
$e^{i\alpha}\boldsymbol{\Psi}(z)$ is also a solution for any real $\alpha$. 
As a consequence, eigenfunctions obtained from independent calculations cannot be
compared directly without fixing a common gauge.

To align local and global eigenfunctions, a single complex scalar $C$ is determined
by minimising the squared difference between the electrostatic potentials over the
central ballooning region,
\begin{equation}
C
=
\arg\min_{c\in\mathbb{C}}
\int_{|z|\le z_0}
\left|
\phi_{\mathrm{glob}}(z) - c\,\phi_{\mathrm{loc}}(z)
\right|^2
\,\mathrm{d}z .
\end{equation}
The minimiser is given analytically by the complex projection
\begin{equation}
C
=
\frac{
\langle \phi_{\mathrm{loc}}, \phi_{\mathrm{glob}} \rangle
}{
\langle \phi_{\mathrm{loc}}, \phi_{\mathrm{loc}} \rangle
},
\qquad
\langle f, g \rangle
\equiv
\int_{|z|\le z_0}
f^{*}(z)\,g(z)\,\mathrm{d}z ,
\end{equation}
where $(\cdot)^{*}$ denotes complex conjugation.

The same complex factor $C$ is applied uniformly to all components of the local
eigenvector, $(\phi, A_{\parallel}, B_{\parallel})_{\mathrm{loc}} \longrightarrow C\,(\phi, A_{\parallel}, B_{\parallel})_{\mathrm{loc}}$,
thereby preserving the physically meaningful relative phases between the fields.
After phase alignment, an overall sign is fixed by requiring the real part of
$\phi$ at its dominant peak to be positive. For shape comparisons, the
electrostatic potential of each calculation is normalised by
$\max_{|z|\le z_0} |\mathrm{Re}(\phi)|$, with the same factor applied to
$A_{\parallel}$ and $B_{\parallel}$ in order to preserve the electromagnetic
polarisation of the mode.

\section{Numerical sources, profile pinning and relaxation} \label{app:sources}

The interior source model has two components, which act like the proportional
and integral terms of a feedback controller. The primary component is a pair of
Krook-type operators with fixed rates $c_{k,\mathrm{heat}}$ and
$c_{k,\mathrm{part}}$, which relax the flux-surface-averaged,
velocity-space-even part of the perturbed distribution towards its target. The
two channels are constructed to be independent: the heat source is projected so
as to inject no net particles, and the energy carried by the particle source is
compensated in the heat channel. The secondary component is a weak integral
corrector: the residual profile deviation is accumulated with an exponential
memory time of $10\,L_{\mathrm{ref}}/c_{\mathrm{ref}}$ and fed back with gains
of $10^{-2}$ in the heat channel and $5\times10^{-3}$ in the particle channel,
removing the finite offset at which a purely proportional term would
equilibrate. Both source profiles are smoothed radially so that no sharp
source layers are driven, and neither component acts on the fluctuating part of
$\delta\!f_\sigma$, so the sources cannot inject fluctuations directly. The
proportional rates are held at
$c_{k,\mathrm{heat}} = c_{k,\mathrm{part}} = 0.01\,c_{\mathrm{ref}}/L_{\mathrm{ref}}$
throughout, for the reasons set out in
Section~\ref{subsec:boundary_conditions_numerical_sources}.

The measurements below use the nominal-$\beta_e$ simulation of
Section~\ref{subsec:nonlinear_GK_STEP-EC-HD}. At the scale of the profiles
themselves, the target, the time average over the quasi-stationary phase, and
the final state are indistinguishable (upper panels of
Figure~\ref{fig:evolved_profiles_multitime}); the deviation panels drawn
beneath them show where the differences lie. The time-averaged electron
temperature departs from its target by at most $0.17$\% over
$0.35 \le x/a \le 0.65$, the ion temperature by $2.0$\%, and the electron
density by $0.21$\%.

The sources pin the profiles on average but not instantaneously: the
instantaneous deviations reach $1.4$\% in $T_e$ and $1.0$\% in $n_e$ during
the strongest transport burst ($t \approx 208\,L_{\mathrm{ref}}/c_{\mathrm{ref}}$),
and are concentrated in the strongly driven outer half of the domain. Despite the very large transient fluxes
the profiles therefore remain gradient driven in the intended sense.

\begin{figure}
  \centering
  \includegraphics[]{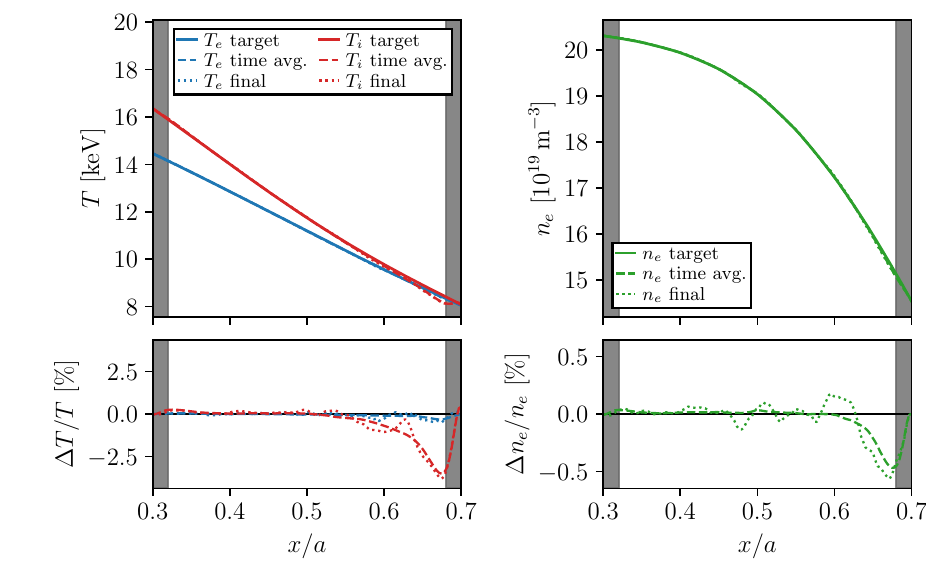}
  \caption{Evolved profiles of the nominal-$\beta_e$ global simulation: electron and ion temperatures (left-hand side) and electron density (right-hand side), showing for each the target, the time average over the quasi-stationary phase, and the final state, with the deviation of the time average and of the final state from the target drawn beneath each panel. Grey shading marks the buffer regions.}
  \label{fig:evolved_profiles_multitime}
\end{figure}

\subsection*{Source-free relaxation}

Every flux reported in Section~\ref{sec:nonlinear} is obtained with the
flux-surface-averaged profiles held close to their targets by the numerical
sources, and a source strong enough to hold the profiles in place might also
be strong enough to sustain the transport it holds them against. To test this
directly, we continue the nominal simulation with both source rates set to
zero, leaving the profiles free to evolve under their own turbulent transport;
Figure~\ref{fig:relaxation} follows the continuation.

The turbulence survives the removal of the sources and then decays with the
profiles that drive it, but it does not decay indefinitely. The total heat
flux falls from its saturated value by roughly an order of magnitude and is
still falling slowly at the end of the simulation; the late-time flux is
carried almost entirely by the electron electromagnetic channel, while the
electron electrostatic, ion electrostatic and ion electromagnetic channels
decay throughout. The
deviation of the electron temperature from its target grows from well below
one per cent when the sources are removed to around six per cent; the
transport at this operating point is therefore very stiff, with a profile
deviation of only a few per cent enough to change the flux by an order of
magnitude. The relaxation itself follows the transport. The electron
temperature profile pivots rather than subsiding uniformly, falling inside
mid-radius and rising outside, so that heat is redistributed from the inner
half of the domain to the outer half, and the relaxation begins in the
strongly driven outer half, where the flux is largest, before moving inward
as the outer gradients degrade and the local drive weakens.

The relaxation slows as it proceeds, and part of this slowing is imposed
rather than physical. The simulated annulus is bounded and the profiles are
held at its edges, so the temperature drop across the domain is fixed: the
interior can redistribute its gradients towards the boundaries but cannot
dispose of them. The $\delta\!f$ ordering also ceases to be well satisfied as
the evolved profiles depart from the fixed background on which the splitting
is built; the profile deviation stays at the few-per-cent level, but because
it is radially localised the corresponding changes in the driving gradients
are much larger, and the ordering assumes that the evolved state stays close
to the background that supplies those drives.

The magnetic geometry is not re-solved as the profiles evolve, so the
$\beta^\prime$ stabilisation~\cite{bourdelle2003} that enables the
flux-driven steady state of~\cite{giacomin2024b} is not available to these
simulations. In bootstrap-current dominated plasmas such as STEP the current
profile and the equilibrium are largely determined by the pressure profile
they confine, and should be recalculated self-consistently as the profiles
evolve. The final state of Figure~\ref{fig:relaxation} is therefore not a
prediction of where a flux-driven STEP-EC-HD would settle. Following the
relaxation further, or all the way to a genuine flux-driven steady state,
requires coupling gyrokinetics to a transport solver that evolves the
profiles and the equilibrium on the transport time scale, as in the
GENE-TANGO~\cite{disiena2022} and CGYRO-PORTALS~\cite{rodriguez2022}
frameworks, and lies beyond the scope of the present work.

\begin{figure}
  \centering
  \begin{minipage}[t]{0.49\textwidth}
    \centering
    \includegraphics[width=\textwidth]{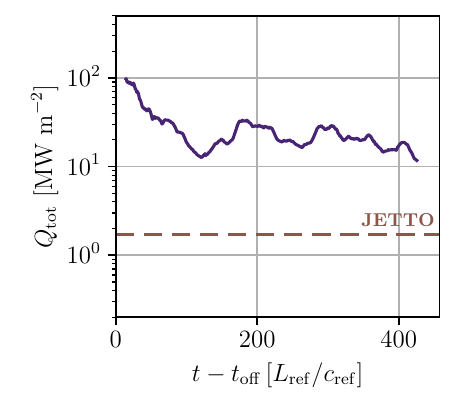}
  \end{minipage}\hfill
  \begin{minipage}[t]{0.49\textwidth}
    \centering
    \includegraphics[width=\textwidth]{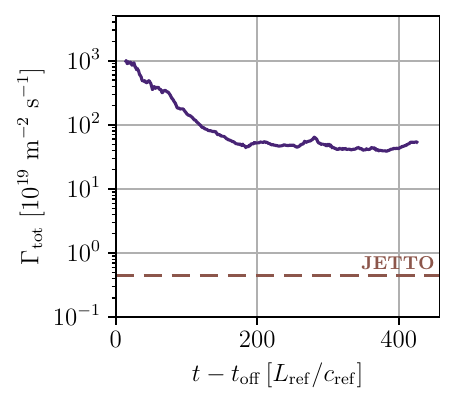}
  \end{minipage}

  \vspace{1em}
  \begin{minipage}[t]{0.49\textwidth}
    \centering
    \includegraphics[width=\textwidth]{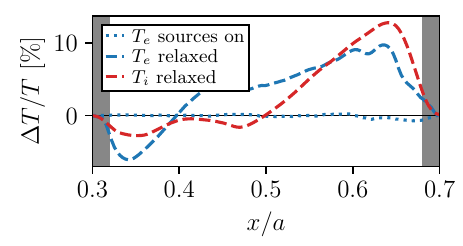}
  \end{minipage}\hfill
  \begin{minipage}[t]{0.49\textwidth}
    \centering
    \includegraphics[width=\textwidth]{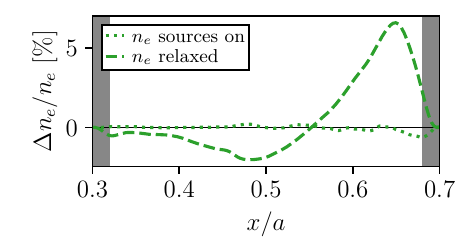}
  \end{minipage}
  \caption{Continuation of the nominal-$\beta_e$ global simulation (G) with the numerical sources switched off at $t_{\mathrm{off}} = 442.2\,L_{\mathrm{ref}}/c_{\mathrm{ref}}$. Top: total heat (left-hand side) and particle (right-hand side) fluxes of the source-free continuation against time from the switch-off, each the volume average over the full radial domain, not the value at any one radius. Brown dashed lines mark balance with the available sources, computed from the SPR-045 JETTO run underlying~\cite{tholerus2024} as in Table~\ref{tab:flux_comparison} (the sources integrated within each flux surface and divided by its area): the panels use the mid-radius values, $1.7\,\mathrm{MW\,m^{-2}}$ and $0.45\times10^{19}\,\mathrm{m^{-2}\,s^{-1}}$. Bottom: deviation of the temperatures (left-hand side) and of the electron density (right-hand side) from their targets, at the end of the simulation with sources (dotted) and of the source-free continuation (dashed). Grey shading marks the buffer regions.}
  \label{fig:relaxation}
\end{figure}

\subsection*{Sensitivity to the source strength}

The strength of the profile control was varied directly, repeating the nominal
case with $c_k$ reduced from $0.01$ to $0.0075$, $0.005$, $0.002$ and $0.001$,
and to zero in the source-free continuation above. Measured as the root mean
square of $|\Delta T_e|/T_e$ over $0.35 \le x/a \le 0.65$, the deviation
averages $0.058$\% over the quasi-stationary phase at the nominal strength and
reaches $6.2$\% with the sources removed. The intermediate cases were started
from noise and are too short to separate from one another, and establish only
that weakening the sources by an order of magnitude does not release the
profile. The high-flux state of Section~\ref{sec:nonlinear} is therefore a property of
the design gradients rather than of the strength chosen for the profile
control.

\section{Reduced-$\beta_e$ comparison at fixed geometry} \label{app:lowbeta}

In this comparison, $\beta_e$ is lowered at fixed magnetic geometry. We compare the nominal case of Section~\ref{subsec:nonlinear_GK_STEP-EC-HD} with one in which the density profile is scaled by~$1/6$, chosen to bring~$\beta_e$ at the centre of the global domain below the threshold identified in~\cite{Kennedy2025EM}. Multiplying the density profile by a constant leaves every logarithmic gradient, and every normalised instability drive, unchanged. Both simulations use the same numerical grid.

This experiment is deliberately simple, and three limitations qualify the statements below. First, the Grad--Shafranov equation is not re-solved, so the reduced-$\beta_e$ case is not in magnetohydrodynamic force balance and retains the Shafranov shift and local-shear stabilisation of the nominal equilibrium\footnote{There is no unique way to re-solve the equilibrium at reduced pressure. In STEP, where the majority of the plasma current is carried as bootstrap current, the current profile, and hence the equilibrium, is largely determined by the pressure profile it confines. Holding the geometry fixed gives a controlled experiment, in which any recovery of zonal-flow regulation at reduced $\beta_e$ can be attributed to the electromagnetic terms in the gyrokinetic equations.}. Second, the equilibrium pressure-gradient term entering the drifts, $\beta^\prime$, is held at its nominal value, so the case is not a fully self-consistent $\beta$ scan (such as are performed in~\cite{Kennedy2025EM}). Third, the density is scaled at fixed temperature, so the collisionality falls by the same factor of six and the comparison is not at fixed collisionality. Subject to these caveats, the experiment isolates the effect of the kinetic pressure entering the gyrokinetic equations.

\begin{figure}
  \centering
  \begin{minipage}[t]{0.49\textwidth}
    \centering
    \includegraphics[width=\textwidth]{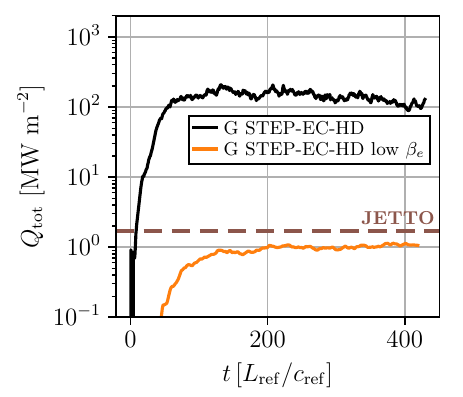}
  \end{minipage}\hfill
  \begin{minipage}[t]{0.49\textwidth}
    \centering
    \includegraphics[width=\textwidth]{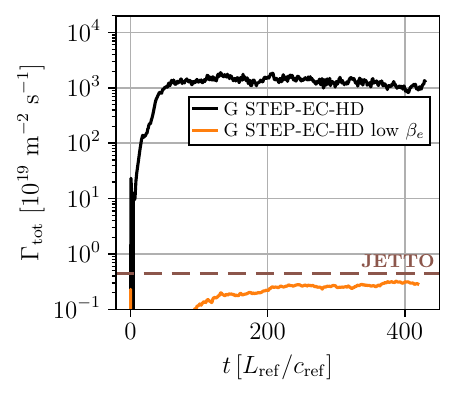}
\end{minipage}
  \caption{Time traces of the volume-averaged total heat (left-hand side) and particle (right-hand side) fluxes from global nonlinear simulations of STEP-EC-HD. Results are shown for two simulations: (i) the nominal kinetic profiles; and (ii) a case where $\beta_e$ is lowered by a factor of six by scaling the density profile while the underlying Grad--Shafranov equilibrium (and hence the equilibrium pressure-gradient term entering the drifts) is held fixed. Brown dashed lines mark balance with the available sources, computed from the SPR-045 JETTO run underlying~\cite{tholerus2024} as in Table~\ref{tab:flux_comparison} (the sources integrated within each flux surface and divided by its area): the panels use the mid-radius values, $1.7\,\mathrm{MW\,m^{-2}}$ and $0.45\times10^{19}\,\mathrm{m^{-2}\,s^{-1}}$.
}
  \label{fig:global_nonlinear_comparison_lowbeta}
\end{figure}

At nominal $\beta_e$ the total heat and particle fluxes of Figure~\ref{fig:global_nonlinear_comparison_lowbeta} saturate at very large values characteristic of the high-flux regime of Section~\ref{subsec:nonlinear_GK_STEP-EC-HD}, with a late-time electron heat flux of order $100\,\mathrm{MW\,m^{-2}}$. At reduced $\beta_e$ the system saturates instead in a zonally regulated state with an electron heat flux of approximately $1\,\mathrm{MW\,m^{-2}}$, sustained for around $280\,L_{\mathrm{ref}}/c_{\mathrm{ref}}$, and a particle flux smaller by orders of magnitude. The two cases therefore bracket the electromagnetic transition, and locate it consistently with the local threshold of~\cite{Kennedy2025EM}: in the local convention of that reference the transition for this equilibrium sits at $q^2\beta_e \approx 0.3$ ($\beta_e \approx 0.026$ at $q = 3.5$ on the $\rho_{\mathrm{tor}} = 0.55$ surface), the nominal case sits at $q^2\beta_e \approx 1.1$, and the sixfold reduction brings the reduced-$\beta_e$ case to $q^2\beta_e \approx 0.18$, below the threshold.

The electron channel carries two thirds of the heat flux at nominal $\beta_e$ and essentially all of it at reduced $\beta_e$, where the ion heat flux falls to $Q_i \approx 0.01\,\mathrm{MW\,m^{-2}}$; the collisionality, which falls with the density by the same factor of six, may also strengthen the trapped-electron drive in this case. Transport at reduced $\beta_e$ nevertheless remains electromagnetic.

\begin{figure}
  \centering
  \includegraphics[]{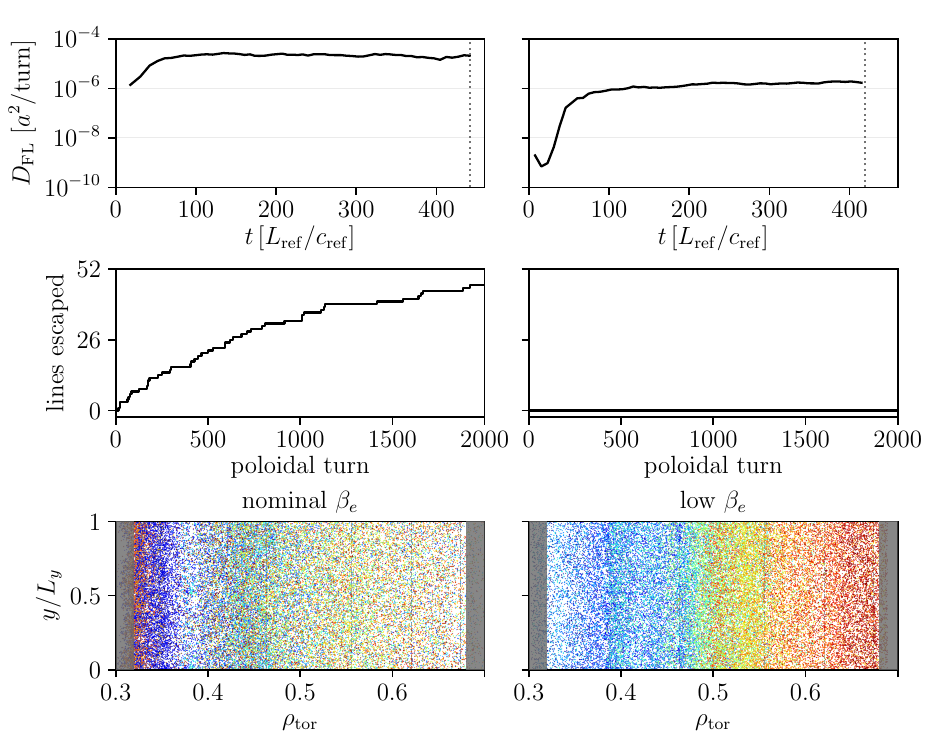}
  \caption{Magnetic stochasticity in the nominal-$\beta_e$ (left-hand side) and reduced-$\beta_e$ (right-hand side) saturated states of the global simulations (G). Top: field-line diffusivity $D_{\mathrm{FL}}$ against time, with the grey dotted line indicating times corresponding to the Poincar\'e plots below. Middle: the number of the $52$ traced field lines that have left the domain, against poloidal turn number. Bottom: Poincar\'e sections at $t=442$ and $419\,L_{\mathrm{ref}}/c_{\mathrm{ref}}$, the lines traced for $2000$ poloidal turns through $\delta B_\perp = \nabla A_{1,\parallel} \times \hat{b}$, one colour each. Dashed lines mark the surfaces $q = m/2$, and grey shading the buffer regions.}
  \label{fig:poincare_beta}
\end{figure}

At nominal $\beta_e$ the Poincar\'e sections of Figure~\ref{fig:poincare_beta} show the confining flux surfaces destroyed across the entire domain, with no island chains surviving at the low-order rational surfaces and $46$ of the $52$ traced field lines leaving the domain, the first within $16$ poloidal turns. At reduced $\beta_e$ the field remains chaotic locally, but the radial excursions are bounded, no traced line leaves the domain within $2000$ turns, and partial transport barriers survive. The field-line diffusivity falls by a factor of about $16$ between the two cases, matching the ratio of the saturated electron heat fluxes in each case's own gyro-Bohm units. Increased magnetic-field-line stochasticisation appears clearly to be associated with the transition to the high-flux regime.

\begin{figure}
  \centering
  \begin{minipage}[t]{0.5\textwidth}
    \centering
    \includegraphics[]{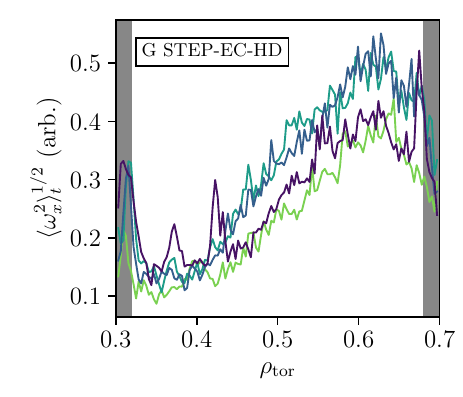}
  \end{minipage}\hfill
  \begin{minipage}[t]{0.5\textwidth}
    \centering
    \includegraphics[]{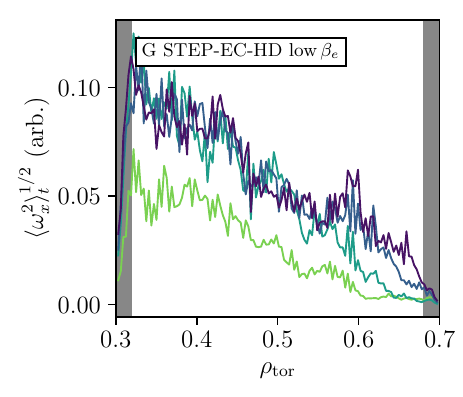}
\end{minipage} \\ \begin{minipage}[t]{0.5\textwidth}
    \centering
    \includegraphics[]{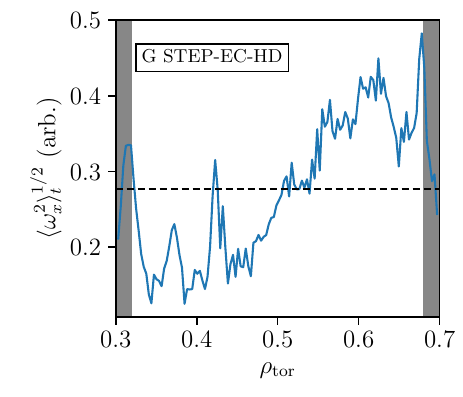}
  \end{minipage}\hfill
  \begin{minipage}[t]{0.5\textwidth}
    \centering
    \includegraphics[]{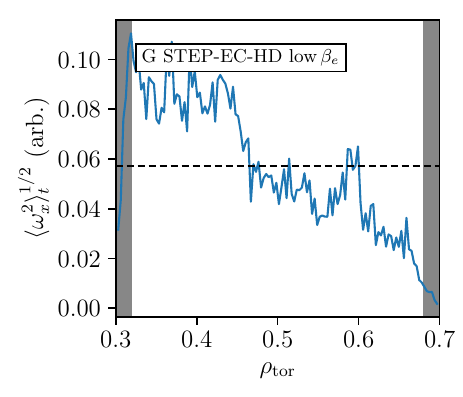}
\end{minipage}

  \caption{Radial structure of the zonal flows in the nonlinear global simulations at nominal $\beta_e$ (left-hand side) and reduced $\beta_e$ (right-hand side). Top: radial profiles of the root-mean-square zonal shearing rate $\omega_x$ over successive quarters of each simulation period, the curves darkening with time. Bottom: the same measure over the final $20$\% of each simulation, dashed black lines the radial averages. Grey shading marks the buffer regions.}
  \label{fig:zonal_evolution}
\end{figure}

Figure~\ref{fig:zonal_evolution} shows radial profiles of the root-mean-square zonal shearing rate $\omega_x$ in the two simulations, defined in Equation~(\ref{eq:zonal_flow}) below. At nominal $\beta_e$ the zonal shear is larger in absolute terms, roughly a factor of five above the reduced-$\beta_e$ level, and grows towards the strongly driven outer region. Electromagnetic stresses oppose zonal-flow generation at nominal $\beta_e$~\cite{Zhang2025a,Zhang2025b,Kennedy2025EM}, and the stochastic field of Figure~\ref{fig:poincare_beta} opens an electron transport channel along the perturbed field lines that acts to weaken the zonal flows. The zonal flow has a magnetic counterpart, the zonal field, which is examined below.

The reduced-$\beta_e$ simulation is qualitatively consistent with the global Cyclone-Base-Case-like studies of~\cite{ishizawa2019,masui2022}, in which zonal flows maintain a steady saturated state; the differences that separate those studies from STEP-EC-HD are discussed in the conclusions.

\subsection*{Zonal flow and zonal field}

Both the zonal flow and the zonal field enter the torque-competition picture of~\cite{Zhang2025a,Zhang2025b}, so we measure each on the two sides of the $\beta_e$ bracket. We write the zonal part of a fluctuating field as its flux-surface average, the component with toroidal and poloidal mode numbers $n = m = 0$,
\begin{equation}
  \bar{f}(x,t) \;=\; \langle f \rangle_{y,z} \;=\;
    \frac{\int \mathcal{J}\, f \,\mathrm{d}y\,\mathrm{d}z}{\int \mathcal{J} \,\mathrm{d}y\,\mathrm{d}z} ,
  \label{eq:zonal_average}
\end{equation}
with $\mathcal{J}$ the Jacobian of the field-aligned coordinates, so that $\bar{\phi}$ and $\bar{A}_{1,\parallel}$ are the zonal potential and the zonal field. The zonal potential drives a binormal $E\times B$ flow and the radial shear of that flow,
\begin{equation}
  v_y \;=\; \frac{\partial \bar{\phi}}{\partial x} ,
  \qquad
  \omega_x \;=\; -\,\frac{\partial^{2} \bar{\phi}}{\partial x^{2}} \;=\; -\,\frac{\partial v_y}{\partial x} ,
  \label{eq:zonal_flow}
\end{equation}
which is the quantity plotted in Figure~\ref{fig:zonal_evolution}.

The zonal field is measured through the corrugation it imposes on the field-line pitch, which is dimensionless and so comparable across the $\beta_e$ bracket,
\begin{equation}
  \delta q \;=\; \frac{k_{y,\min}}{n_{0}^{\mathrm{min}}}
    \left[\, \Lambda\,\frac{\partial \langle A_{1,\parallel} \rangle_{y}}{\partial \rho_{\mathrm{tor}}} \,\right]_{z} ,
  \qquad
  \delta \hat{s} \;=\; \frac{\rho_{\mathrm{tor}}}{q}\,\frac{\partial\, \delta q}{\partial \rho_{\mathrm{tor}}} ,
  \label{eq:dshat}
\end{equation}
where $\langle \cdot \rangle_{y}$ is the average over $y$ alone and $[\, \cdot \,]_{z}$ the unweighted mean along the field line, taken after the product is formed, as the field-line equations require. Here $\Lambda = \mathcal{J}\,|\nabla x \times \nabla y|^{2}/B_{0}$ is the metric factor relating $A_{1,\parallel}$ to the field-line pitch, $k_{y,\min}$ the smallest finite binormal wavenumber on the grid and $n_{0}^{\mathrm{min}}$ its toroidal mode number. Finally, $\delta\hat{s}$ is the corrugation the zonal field imposes on the equilibrium magnetic shear $\hat{s} = (\rho_{\mathrm{tor}}/q)\,\partial q/\partial \rho_{\mathrm{tor}}$.

\begin{figure}
  \centering
  \begin{minipage}[t]{0.5\textwidth}
    \centering
    \includegraphics[]{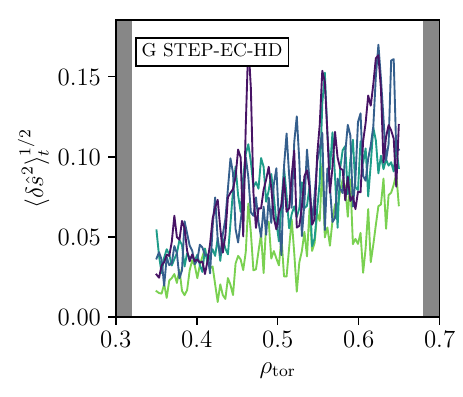}
  \end{minipage}\hfill
  \begin{minipage}[t]{0.5\textwidth}
    \centering
    \includegraphics[]{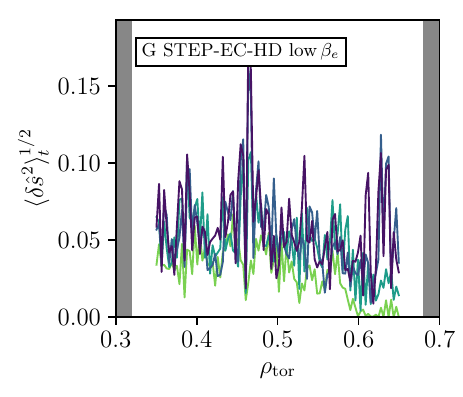}
\end{minipage} \\ \begin{minipage}[t]{0.5\textwidth}
    \centering
    \includegraphics[]{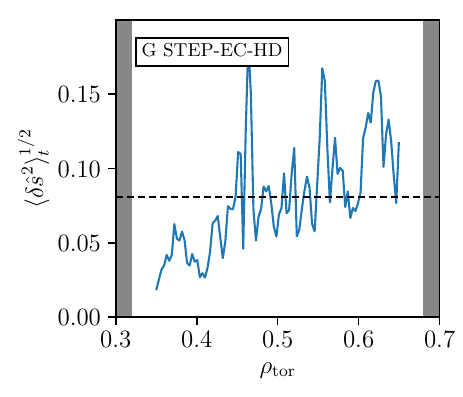}
  \end{minipage}\hfill
  \begin{minipage}[t]{0.5\textwidth}
    \centering
    \includegraphics[]{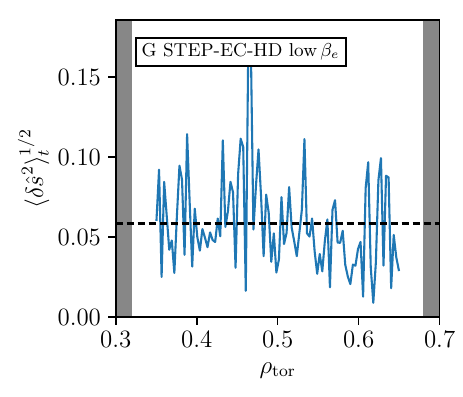}
\end{minipage}

  \caption{Radial structure of the zonal field at nominal $\beta_e$ (left-hand side) and reduced $\beta_e$ (right-hand side). Top: root-mean-square corrugation of the magnetic shear $\delta\hat{s}$ over successive quarters of each simulation period, the curves darkening with time. Bottom: the same measure over the final $20$\%, dashed black lines the radial averages. Grey shading marks the buffer regions.}
  \label{fig:zonal_field}
\end{figure}

Figure~\ref{fig:zonal_field} shows that the zonal field weakens far less across the bracket than the zonal flow. Averaged over the final $20$\% of each simulation and over $0.35 \le x/a \le 0.65$ (the dashed lines of Figure~\ref{fig:zonal_field}), the rms corrugation of the magnetic shear is $0.081$ at nominal $\beta_e$ against $0.059$ at reduced $\beta_e$, and as a fraction of the perpendicular magnetic energy the zonal field is larger at reduced $\beta_e$ ($0.16$ against $0.078$). The reduced-$\beta_e$ state therefore retains a strong zonal field despite its much lower zonal flow shear (Figure~\ref{fig:zonal_evolution}).

\section{Equilibrium flow shear} \label{app:flowshear}

\begin{figure}
  \centering
  \begin{minipage}[t]{0.49\textwidth}
    \centering
    \includegraphics[width=\textwidth]{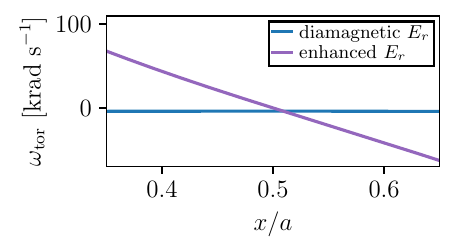}
  \end{minipage}\hfill
  \begin{minipage}[t]{0.49\textwidth}
    \centering
    \includegraphics[width=\textwidth]{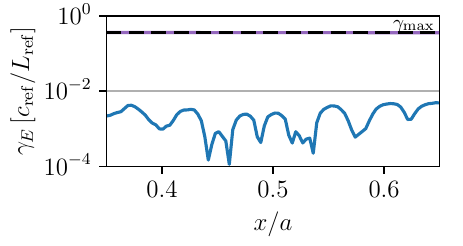}
  \end{minipage}
  \caption{Imposed toroidal rotation profiles (left-hand side) and the corresponding $E\times B$ shearing rates (\ref{eq:gammaE}) (right-hand side), for the two global simulations with an equilibrium radial electric field. The diamagnetic case is nearly a rigid rotation; the other is a linear ramp giving radially uniform $\gamma_E \approx \gamma_{\mathrm{max}}$, the black dashed line marking $\gamma_{\mathrm{max}} = 0.36\,c_{\mathrm{ref}}/L_{\mathrm{ref}}$. Only the interior $0.35 \le x/a \le 0.65$ of the domain is shown.}
  \label{fig:flowshear_profiles}
\end{figure}

The nonlinear simulations of Section~\ref{sec:nonlinear} carry no equilibrium
radial electric field. We therefore repeat the nominal case with two imposed
rotation profiles, entering as an equilibrium radial electric field through the
force balance of (\ref{eq:force_balance}) and shown with their shearing
rates in Figure~\ref{fig:flowshear_profiles}: (i) one reproducing the
diamagnetic field of Section~\ref{subsec:numerical_setup}, whose mean shearing
rate is two orders of magnitude below $\gamma_{\mathrm{max}}$; and (ii) a
synthetic linear ramp giving radially uniform
$\gamma_E \approx \gamma_{\mathrm{max}}$. We note that STEP-EC-HD has no external source of torque to drive such rotation.

\begin{figure}
  \centering
  \begin{minipage}[t]{0.49\textwidth}
    \centering
    \includegraphics[width=\textwidth]{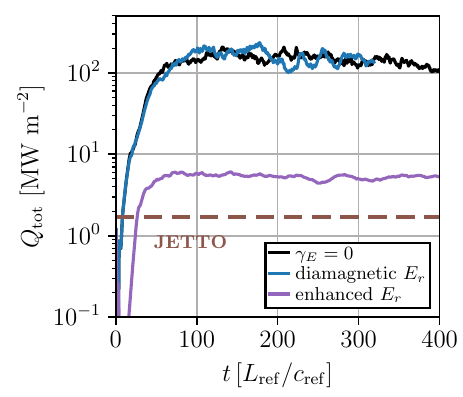}
  \end{minipage}\hfill
  \begin{minipage}[t]{0.49\textwidth}
    \centering
    \includegraphics[width=\textwidth]{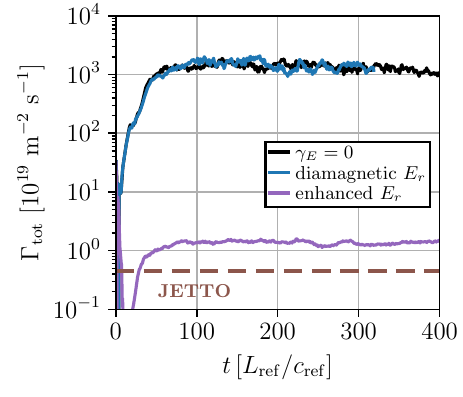}
  \end{minipage}

  \vspace{1em}
  \begin{minipage}[t]{0.49\textwidth}
    \centering
    \includegraphics[width=\textwidth]{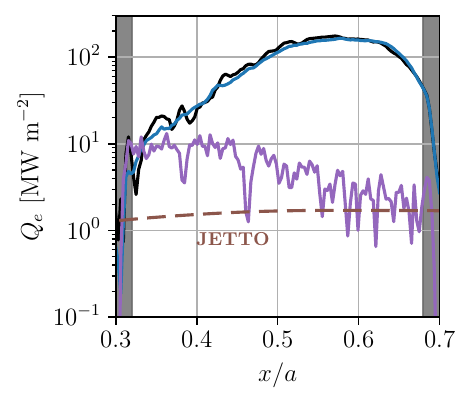}
  \end{minipage}\hfill
  \begin{minipage}[t]{0.49\textwidth}
    \centering
    \includegraphics[width=\textwidth]{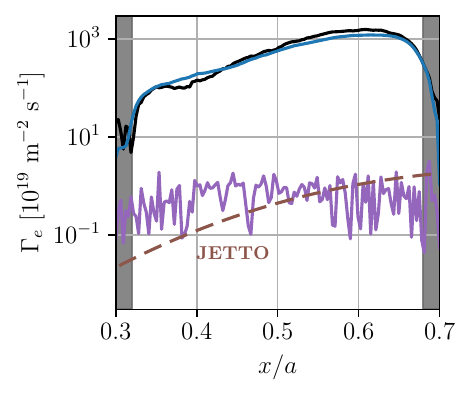}
  \end{minipage}
  \caption{Global nonlinear simulations without equilibrium flow shear and with the two rotation profiles of Figure~\ref{fig:flowshear_profiles}. Top: time traces of the volume-averaged total heat (left-hand side) and particle (right-hand side) fluxes. Bottom: radial profiles of the electron heat (left-hand side) and particle (right-hand side) fluxes, averaged over the final quarter of each simulation. Brown dashed lines mark balance with the available sources, computed from the SPR-045 JETTO run underlying~\cite{tholerus2024} as in Table~\ref{tab:flux_comparison} (the sources integrated within each flux surface and divided by its area): the top panels use the mid-radius values, $1.7\,\mathrm{MW\,m^{-2}}$ and $0.45\times10^{19}\,\mathrm{m^{-2}\,s^{-1}}$, and the bottom panels the radially resolved profiles, which vary with radius. Grey shading marks the buffer regions.}
  \label{fig:flowshear_fluxes}
\end{figure}

At the diamagnetic level, flow shear has no discernible effect, with the domain-averaged heat flux in Figure~\ref{fig:flowshear_fluxes} remaining indistinguishable from the $\gamma_E=0$ baseline. At the quench-rule level, however, the heat flux is reduced by more than an order of magnitude and the particle flux by nearly three orders of magnitude, with the residual transport dominated by the electromagnetic electron heat flux. Both fluxes are consequently brought substantially closer to the levels compatible with the available sources: the particle flux is approximately balanced by the available fuelling, while the heat flux remains only several times larger than the available heating power. Although none of the global simulations considered here is fully quenched by the imposed flow shear, these results suggest that sufficiently strong flow shear may provide a route for STEP to access a more favourable transport regime.

A finer scan has been performed within the local framework for this operating point, as reported in \cite{giacomin2023b} (their Figure 3). In their simulations fluxes fall below the source-compatible level for $\gamma_E = 0.05$–$0.1\,c_s/a$ which is comparable to $\gamma_{\mathrm{max}}$ on the $\rho_{\rm tor} = 0.55$ surface considered there (Figure 2 of \cite{giacomin2023b}).

\bibliographystyle{unsrt}
\bibliography{bibliography}

\end{document}